\documentclass[pdflatex,sn-basic,a4paper]{sn-jnl}

\usepackage{manyfoot}
\usepackage[dvipsnames]{xcolor}
\usepackage{amsmath}

\usepackage{cite}
\usepackage{algorithmic}
\usepackage{graphicx}
\usepackage{textcomp}
\usepackage{orcidlink}
\usepackage{enumerate}

\usepackage[T1]{fontenc}  

\usepackage{amsthm}
\usepackage{amsmath}
\usepackage{amssymb}
\usepackage{amsfonts}

\usepackage{wrapfig}  
\usepackage{float}  

\PassOptionsToPackage{svgnames,table}{xcolor}
\definecolor{darkblue}{rgb}{0, 0, 0.7}

\usepackage[capitalise,noabbrev]{cleveref}

\crefname{lstlisting}{listing}{listings}
\Crefname{lstlisting}{Listing}{Listings}

\usepackage[
disable,
]{todonotes}

\usepackage{paralist}

\usepackage{booktabs}
\usepackage[framemethod=TikZ]{mdframed}
\mdfdefinestyle{rqanswer}{backgroundcolor=gray!10,roundcorner=3pt,outerlinewidth=1}

\usepackage{url}

\usepackage{verbatim}
\usepackage{subcaption}
\usepackage{pifont}
\usepackage[breakable]{tcolorbox}

\definecolor{darkblue}{rgb}{0, 0, 0.7}

\hypersetup{
    colorlinks=true,
    linkcolor=darkblue,
    filecolor=darkblue,      
    urlcolor=blue!50!blue,
    pdftitle={},
    pdfpagemode=FullScreen,
    }

\usepackage[capitalise,noabbrev]{cleveref}

\usepackage{xspace} 

\newcommand{\aka}{a.k.a.\xspace}
\newcommand{\eg}{e.g.\xspace}
\newcommand{\etal}{et al.\xspace}

\newcommand{\ie}{i.e.\xspace}

\definecolor{badred}{RGB}{240,0, 0}
\definecolor{goodgreen}{RGB}{0, 216, 0}
\usepackage{transparent}

\usepackage[acronyms,shortcuts,section=section,style=super,nopostdot,nogroupskip,nomain,nonumberlist]{glossaries}
\glsdisablehyper

\newacronym{at}     {AT}    {actual twin}
\newacronym{cs}     {CS}    {control system}
\newacronym{cps}    {CPS}   {cyber-physical system}
\newacronym{ds}     {DS}    {digital shadow}
\newacronym{dt}     {DT}    {digital twin}
\newacronym{iot}    {IoT}   {internet of things}
\newacronym[longplural={properties of interest}]
    {poi}    {PoI}   {Properties of Interest}

\newcommand{\DT}{\ac{dt}\xspace}
\newcommand{\DTs}{\acp{dt}\xspace}

\newcommand{\AT}{\ac{at}\xspace}
\newcommand{\ATs}{\acp{at}\xspace}

\newcommand{\DarTwin}{\textsf{DarTwin}\xspace}

\usepackage{pgfplots}  

\usepackage{tikz}

\usetikzlibrary{backgrounds}
\usetikzlibrary{positioning}
\usetikzlibrary{calc} 
\usetikzlibrary{arrows}
\usetikzlibrary{arrows.meta}
\usetikzlibrary{shapes}
\usetikzlibrary{shapes.arrows}
\usetikzlibrary{shapes.geometric}
\usetikzlibrary{shapes.symbols}  
\usetikzlibrary{patterns}

\makeatletter
\pgfdeclareshape{umlpentagon}{
  \inheritsavedanchors[from=rectangle] 
  \inheritanchorborder[from=rectangle]
  \inheritanchor[from=rectangle]{center}
  \inheritanchor[from=rectangle]{north}
  \inheritanchor[from=rectangle]{north east}
  \inheritanchor[from=rectangle]{east}
  \inheritanchor[from=rectangle]{south east}
  \inheritanchor[from=rectangle]{south}
  \inheritanchor[from=rectangle]{south west}
  \inheritanchor[from=rectangle]{west}
  \inheritanchor[from=rectangle]{north west}
  \backgroundpath{
    \southwest \pgf@xa=\pgf@x \pgf@ya=\pgf@y
    \northeast \pgf@xb=\pgf@x \pgf@yb=\pgf@y
    \pgf@xc=\pgf@xb \advance\pgf@xc by-5pt 
    \pgf@yc=\pgf@ya \advance\pgf@yc by+5pt
    \pgfpathmoveto{\pgfpoint{\pgf@xa}{\pgf@ya}}
    \pgfpathlineto{\pgfpoint{\pgf@xa}{\pgf@yb}}
    \pgfpathlineto{\pgfpoint{\pgf@xb}{\pgf@yb}}
    \pgfpathlineto{\pgfpoint{\pgf@xb}{\pgf@yc}}
    \pgfpathlineto{\pgfpoint{\pgf@xc}{\pgf@ya}}
    \pgfpathclose
 }
}
\makeatother

\definecolor{darkgreen}{rgb}{0.0, 0.5, 0.0}

\usepackage{tikzpeople}

\tikzset{
    evaluation/.style={-latex,draw=gray},
    circlenum/.style={circle,draw=gray,fill=green!15,yshift=-2pt,inner sep=1.25pt,minimum size=0pt,font=\small\sffamily}
}

\newcommand{\circlenum}[1]{%
\tikz{%
\node[circlenum,font=\small] {#1}
}} %
\usepackage{tikz-uml}
\usepackage{tikz-layers}

\tikzset{
every node/.append style={align=center,font=\sffamily},
block/.append style={align=center,rectangle,draw=black,
minimum height=1cm},
dt/.append style={block,fill=yellow!50},
sys/.append style={block,fill=gray!20,inner sep=2em, inner xsep=.7em,yshift=.75em},
system/.append style={block,fill=gray!20,inner sep=2em},
pattern-at/.append style={align=center,fill=white,rectangle,draw=black,thick,     minimum height=1.5em,minimum width=3em,
    font=\sffamily\fontsize{8}{8}\selectfont},
pattern-dt/.append style={align=center,fill=white,rectangle,draw=black,thick,
    rounded corners=3pt,minimum height=1.5em,minimum width=3em,
    font=\sffamily\fontsize{8}{8}\selectfont},
box/.append style={rectangle,draw=lightgray,thick,inner sep=.25em,
    minimum width=3.25cm,minimum height=1.75cm},
title-box/.append style={rectangle, draw=lightgray, thick, inner sep=.5em,
    minimum width=6.5cm},
patternarrow/.append style={latex-latex,thick},
patternhierarchical/.append style={draw,rectangle,thick,
    inner sep=.75em,inner ysep=1em,
    font=\sffamily\fontsize{8}{8}\selectfont,rounded corners=5pt,
}
}

\definecolor{NiceHighlightColor}{HTML}{FFC107}

\tikzset{
framelabel/.append style={umlpentagon,draw,fill=white,anchor=north west,
    align=left,font=\sffamily\fontsize{8pt}{8pt}\selectfont,
},
dartwinFramed/.style 2 args={
execute at end picture={
    \begin{scope}[on background layer]
      \draw[draw] ($(current bounding box.south west)-(1em,1em)$)
                rectangle ($(current bounding box.north east)+(1em,1.5em)$);
      \node[framelabel,inner xsep=.5em,inner ysep=.25em] at (current bounding box.north west){\textbf{dartwin} #1 \ifstrempty{#2}{}{\textbf{based on} #2}};
    \end{scope}
}},
portlabel/.append style={align=center,font=\sffamily\fontsize{8pt}{8pt}\selectfont},
langElement/.append style={draw=black,fill=white,font=\sffamily\scriptsize},
influence/.append style={langElement, signal, signal to=east and west,inner ysep=.75em},
sensor/.append style={langElement, trapezium, 
    trapezium left angle=75, trapezium right angle=105,
    minimum height=2em,inner ysep=.5em, inner sep=.5em},
actuator/.append style={langElement, trapezium, 
    trapezium left angle=105, trapezium right angle=75,
    minimum height=2em,inner ysep=.5em, inner sep=.5em},
process/.append style={langElement, rectangle, rounded corners=.5em,
    inner sep=1em,minimum width=3cm},
arbiter/.append style={langElement,shape=regular polygon, regular polygon sides=8,inner sep=0em},
actualtwin/.append style={langElement, rectangle, inner sep=1em,minimum width=5em},
poi/.append style={langElement, ellipse,minimum height=3em, minimum width=5em},
cvar/.append style={poi},
mvar/.append style={poi},
cmvar/.append style={poi},
evar/.append style={poi},
goal/.append style={langElement, trapezium, 
    trapezium left angle=75, trapezium right angle=75, 
    inner ysep=.75em, minimum height=3em},
langDT/.append style={langElement,fill=none,dashed,thick,inner sep=1em, inner ysep=1.5em,yshift=.7em},
separationLine/.append style={draw,ultra thick},
label/.append style={font=\large\sffamily\bfseries},
fullarrow/.append style={-latex,thick},
dashedarrow/.append style={-latex,thick,dashed},
arrowlabel/.append style={sloped,below,font=\sffamily\scriptsize},
specialization/.append style={-{Triangle[open,length=3mm, width=3mm]},thick},
contributes/.append style={-{Straight Barb[length=3mm, width=3mm]},thick},
relation/.append style={-{Stealth[length=3mm, width=3mm]},dashed,thick},
flow/.append style={-{Stealth[length=2mm, width=2mm]},thick},
conflict/.append style={{Triangle[length=3mm, width=3mm]}-{Triangle[length=3mm, width=3mm]},thick,double},
goalrelation/.append style={{Triangle[length=3mm, width=3mm]}-{Triangle[length=3mm, width=3mm]},thick},
measured/.append style={fullarrow},
determined/.append style={dashedarrow},
implemented/.append style={dashedarrow},
consumed/.append style={dashedarrow},
derived/.append style={dashedarrow,dotted},
controlled/.append style={dashedarrow},
highlight/.append style={draw=NiceHighlightColor,very thick},
highlightnode/.append style={draw=NiceHighlightColor,very thick,fill=NiceHighlightColor!20},
port/.append style={draw=black,fill=white,font=\sffamily\fontsize{4}{4}\selectfont,rectangle,inner sep=0em,minimum width=1em,minimum height=1em,
append after command={},
},
pics/separation/.style={code={
    \draw[separationLine] (0,0) -- ++(#1,0);   
}},
pics/frameSeparation/.style={code={
    \draw[separationLine] 
        ($(current bounding box.west |- #1.south)-(0,1.5em)$) -- 
        ($(current bounding box.east |- #1.south)-(0,1.5em)$);
}},
pics/comfort goal/.style={code={
    \node[goal,#1] (comfort goal) {\textbf{Warm Comfort}\\Room Temperature $t$\\ $t$ in range [l,u]};    
}},
pics/freezeprotect goal/.style={code={
    \node[goal,#1] (freezeprotect goal) {\textbf{No Freezing}\\Room Temperature $t$\\ $t$ > 8};    
}},
pics/savemoney goal/.style={code={
    \node[goal,#1] (savemoney goal) {\textbf{Saving Money}\\Energy cost $c$ \\ Lower $c$ by\\ considering temporal\\energy prices};    
}},
pics/energy goals/.style={code={
    \node[goal,#1] (lowerenergy goal) {\textbf{Lower Energy}\\Energy $e$ [kWh]\\lower $e$ than before};
    \node[goal,#1,below right=1 and 1 of lowerenergy goal] (whenpresent goal) {\textbf{Energy when present}\\Energy $e$ [kWh]\\Presence of people $p$\\Same $e$ as before};
    \node[goal,#1,below left=1 and 1 of lowerenergy goal] (whenabsent goal) {\textbf{Energy when absent}\\Energy $e$ [kWh]\\Presence of people $p$\\Lower $e$ than before};

    \draw[specialization,#1] (whenabsent goal.north)  |- ($(whenabsent goal.north)+(0,1em)$) -| (lowerenergy goal);
    \draw[specialization,#1] (whenpresent goal.north) |- ($(whenpresent goal.north)+(0,1em)$) -| (lowerenergy goal);
}},
pics/comfort dt/.style={code={
    \node[process,#1] (-dt) {Thermostat Logic};
    \node[port,#1,label={[portlabel,below]{ }}] (-heater) at (-dt.east) {$\rightarrow$};
    \node[port,#1,label={[portlabel,below]{ }}] (-comfort-temp) at (-dt.west) {$\rightarrow$};
    \node[port,#1,label={[portlabel,below]{ }}] (-temp) at (-dt.south) {$\uparrow$};   
}}, 
pics/energysave dt/.style={code={
    \node[process,#1] (-dt) {Energy Saving};
    \node[port,#1,label={below:{}}] (-comfort-temp-out) at (-dt.east) {$\rightarrow$};
    \node[port,#1,label={below:{}}] (-comfort-temp-in) at (-dt.west) {$\rightarrow$};
    \node[port,#1,label={below:{}}] (-presence) at (-dt.south) {$\uparrow$};   
}}, 
pics/freezeprotect dt/.style={code={
    \node[process,#1] (-dt) {Freeze Protection};
    \node[port,#1,label={below:{}}] (-temp) at (-dt.west) {$\rightarrow$};
    \node[port,#1,label={below:{}}] (-heater) at (-dt.south) {$\downarrow$};   
}}, 
pics/costsave dt/.style={code={
    \node[process,#1] (-dt) {Cost Saving};
    \node[port,#1,label={below:{}}] (-price) at (-dt.west) {$\rightarrow$};
    \node[port,#1,label={below:{}}] (-comfort-temp) at (-dt.east) {$\rightarrow$};   
}}, 
pics/arbiter dt/.style={code={
    \node[arbiter,minimum height=1.5cm,minimum width=1cm,#1] (-dt) {Arbiter};
    \node[port,#1,label={below:{}},rotate=-45] (-in1) at (-dt.north west) {$\rightarrow$};
    \node[port,#1,label={below:{}},rotate=45] (-in2) at (-dt.south west) {$\rightarrow$};
    \node[port,#1,label={below:{}}] (-comfort-temp) at (-dt.east) {$\rightarrow$};   
}}, 
pics/temp heat ports/.style={code={
    \node[port,label={[portlabel]below:{Room Temperature}}] (-room-temp) at (#1.south -| comfort-dt-temp) {$\uparrow$};
    \node[port,label={[portlabel]below:{Heater\\ on/off}},xshift=2em] (-heater) at (#1.south -| comfort-dt-heater) {$\downarrow$};
}},
pics/temp heat ports hooked/.style={code={
    \pic {temp heat ports={#1}};
    \draw[flow] (comfort-dt-heater) -| (-heater);
    \draw[flow] (-room-temp.north) -- (comfort-dt-temp.south);
}},
pics/comfort wrap ports/.style={code={
    \pic {temp heat ports={#1}};
    \node[port,label={[portlabel]left:{User Comfort\\Temperature}}] (-comfort-temp) at (#1.west |- comfort-dt-comfort-temp) {$\rightarrow$};
}},
pics/comfort wrap ports-highlight/.style={code={
    \pic {temp heat ports={#1}};
    \node[port,highlight,label={[portlabel]left:{User Comfort\\Temperature}}] (-comfort-temp) at (#1.west |- comfort-dt-comfort-temp) {$\rightarrow$};
}},
pics/comfort wrap ports nolabel/.style={code={
    \node[port,label={below:{}}] (-room-temp) at (#1.south -| comfort-dt-temp) {$\uparrow$};
    \node[port,label={below:{}},xshift=2em] (-heater) at (#1.south -| comfort-dt-heater) {$\downarrow$};
    \node[port,label={above:{}}] (-comfort-temp) at (#1.west |- comfort-dt-comfort-temp) {$\rightarrow$};
}},
pics/comfort wrap ports hooked/.style={code={
    \pic {comfort wrap ports={#1}};
    \draw[flow] (comfort-dt-heater) -| (-heater);
    \draw[flow] (-room-temp.north) -- (comfort-dt-temp.south);
}},
pics/comfort wrap ports hooked-highlight/.style={code={
    \pic {comfort wrap ports-highlight={#1}};
    \draw[flow] (comfort-dt-heater) -| (-heater);
    \draw[flow] (-room-temp.north) -- (comfort-dt-temp.south);
}},
pics/refactored warm comfort AT/.style={code={
    \begin{scope}[on behind layer]
       \node[sys,fill=gray!10,label={[below]north:{\textbf{twin system} Warm Comfort as AT}},minimum width=12cm,minimum height=3.5cm] (-system) at (-2.5,-1) {};
    \end{scope}
    \node[port,label={[portlabel]below:{Room Temperature}}] (-room-temp) at (-system.south -| comfort-dt-temp) {$\uparrow$};
    \node[port,label={[portlabel]below:{Heater I/O}},xshift=2em] (-heater) at (-system.south -| comfort-dt-heater) {$\downarrow$};
    \node[port,label={[portlabel]above:{User Comfort Temperature}}] (-comfort-temp) at (-comfort-dt-comfort-temp -| -system.west) {$\rightarrow$};
    \draw[flow] (-comfort-dt-heater) -| (-heater);
    \draw[flow] (-room-temp.north) -- (-comfort-dt-temp.south);
    \draw[flow] (-energysave-dt-comfort-temp-out) |- (-comfort-dt-comfort-temp); 
    \draw[flow] (-comfort-temp) -| ($(-comfort-temp)!.5!(-energysave-dt-comfort-temp-in)$) |- (-energysave-dt-comfort-temp-in); 
}},
}

\tikzset{
pics/schema goal/.style args={#1/#2}{code={
    \node[goal,#2] (-goal) {\textbf{#1}};
}},
pics/schema separation/.style={code={
    \draw[separationLine] (0,0) -- ++(#1,0);   
}},
pics/schema dt only/.style args={#1/#2}{code={
    \node[process,#2,minimum width=2cm] (-dt) {#1};
}},
pics/schema dt/.style args={#1/#2}{code={
    \node[process,#2,minimum width=2cm] (-dt) {#1};
    \pic {schema inout={-dt/#2}};
}},
pics/schema arbiter dt/.style args={#1/#2}{code={
    \node[arbiter,#2,minimum width=1cm] (-dt) {#1};
    \node[port,#2,label={below:{}},rotate=45] (-in1) at (-dt.north west) {$\downarrow$};
    \node[port,#2,label={below:{}},rotate=-45] (-in2) at (-dt.north east) {$\downarrow$};
    \node[port,#2,label={below:{}}] (-out) at (-dt.south) {$\downarrow$};
}},
pics/schema inout/.style args={#1/#2}{code={
    \node[port,#2,label={below:{}},xshift=-1em] (-in) at (#1.south) {$\uparrow$};
    \node[port,#2,label={below:{}},xshift=1em] (-out) at (#1.south) {$\downarrow$};
}},
pics/schema aligned inout/.style args={#1/#2/#3}{code={
    \node[port,#3,label={below:{}}] (-in) at (#1.south -| #2-in) {$\uparrow$};
    \node[port,#3,label={below:{}}] (-out) at (#1.south -| #2-out) {$\downarrow$};
}},
pics/schema aligned inout-highlight/.style args={#1/#2/#3}{code={
    \node[port,#3,label={below:{}}] (-in) at (#1.south -| #2-in) {$\uparrow$};
    \node[port,label={below:{}}] (-out) at (#1.south -| #2-out) {$\downarrow$};
}},
}

\tikzset{
pics/noswing goal/.style={code={
    \node[goal,#1] (noswing goal) {\textbf{No Swinging}\\\textbf{Motion}\\ Angle $\theta$ \\ $\theta_{end\,time} = 0$};    
}},
pics/constraints goal/.style={code={
    \node[goal,#1] (constraints goal) {\textbf{Respect System}\\ \textbf{Constraints}\\ Trace $t$ \\No violations over $t$};    
}},
pics/minimize goal/.style={code={
    \node[goal,#1] (minimize goal) {\textbf{Minimize Trajectory}\\\textbf{Duration}\\ Trace duration $d$ \\ minimize($d$)};    
}},
pics/nocollision goal/.style={code={
    \node[goal,#1] (nocollision goal) {\textbf{Respect Dynamic}\\\textbf{ Position Constraints}\\ Trace $t$\\ No violations over $t$};    
}},
pics/validation goal/.style={code={
    \node[goal,#1] (validation goal) {\textbf{Continuous Validation}\\ Validation metrics $vm$\\$vm$ conform to thresholds};    
}},
pics/container goal/.style={code={
    \node[goal,#1] (container goal) {\textbf{Respect Container}\\\textbf{Kinetic Constraints}\\ Trace $t$\\No violations over $t$};    
}},
pics/trajectory dt/.style={code={
    \node[process,#1,minimum width=4cm] (-dt) {Trajectory};
    \node[port,#1,label={below:{}}] (-position) at ($(-dt.south)!.75!(-dt.south west)$) {$\uparrow$};
    \node[port,#1,label={below:{}}] (-swing) at (-dt.south) {$\uparrow$};
    \node[port,#1,label={below:{}}] (-controllers) at ($(-dt.south)!.75!(-dt.south east)$) {$\downarrow$};   
}},
pics/objectsarea dt/.style={code={
    \node[process,#1,minimum width=4cm] (-dt) {Objects in Area};
    \node[port,#1,label={below:{}}] (-constraints) at (-dt.east) {$\rightarrow$};
    \node[port,#1,label={below:{}}] (-camera) at (-dt.south) {$\uparrow$};   
}},
pics/validation dt/.style={code={
    \node[process,#1,minimum height=1.5cm] (-dt) {Validation};
    \node[port,#1,label={below:{}}] (-controllers) at ($(-dt.west)!.75!(-dt.north west)$) {$\rightarrow$};   
    \node[port,#1,label={below:{}}] (-position) at (-dt.west) {$\rightarrow$};
    \node[port,#1,label={below:{}}] (-swing) at ($(-dt.west)!.75!(-dt.south west)$) {$\rightarrow$};
    \node[port,#1,label={below:{}}] (-metrics) at (-dt.east) {$\rightarrow$};
}},
pics/container dt/.style={code={
    \node[process,#1] (-dt) {Container Specification};
    \node[port,#1,label={below:{}}] (-camera) at (-dt.south) {$\uparrow$};   
    \node[port,#1,label={below:{}}] (-database) at (-dt.west) {$\rightarrow$};
    \node[port,#1,label={below:{}}] (-limits) at (-dt.east) {$\rightarrow$};
}},
pics/gantry arbiter dt/.style={code={
    \node[arbiter,#1,minimum width=1cm] (-dt) {Arbiter};
    \node[port,#1,label={below:{}}] (-north-in) at (-dt.north) {$\downarrow$};
    \node[port,#1,label={below:{}}] (-west-in) at (-dt.west) {$\rightarrow$};
    \node[port,#1,label={below:{}}] (-east-out) at (-dt.east) {$\rightarrow$};   
}},
pics/wrap trajectory ports/.style={code={
    \node[port,label={[portlabel]below:{Motor\\Position}}] (-position) at (#1.south -| trajectory-dt-position) {$\uparrow$};
    \node[port,label={[portlabel]below:{Swing\\Angle}}] (-swing) at (#1.south -| trajectory-dt-swing) {$\uparrow$};
    \node[port,label={[portlabel]below:{Motor\\Controllers}}] (-controllers) at (#1.south -| trajectory-dt-controllers) {$\downarrow$};  
    \draw[flow] (-position) -- (trajectory-dt-position);
    \draw[flow] (-swing) -- (trajectory-dt-swing);
    \draw[flow] (trajectory-dt-controllers) -- (-controllers);
}},
pics/wrap container ports/.style={code={
    \node[port,highlight,label={[portlabel]below:{Container\\Camera}}] (-camera) at (#1.south -| container-dt-camera) {$\uparrow$};
    \node[port,highlight,label={[portlabel]left:{Container\\Database}}] (-database) at (#1.west |- container-dt-database) {$\rightarrow$};
    \draw[flow,highlight] (-camera) -- (container-dt-camera);
    \draw[flow,highlight] (-database) -- (container-dt-database);
}},
}

\newenvironment{revision*}{  
\color{ForestGreen}
}{}

\begin{document}

\title{Continuous Evolution of Digital Twins using the DarTwin Notation
}


\author*[1,4]{\fnm{Joost} \sur{Mertens}{\hypersetup{pdfborder={0 0 0}}\orcidlink{0000-0002-8148-5024}}}\email{joost.mertens@uantwerpen.be}

\author*[2]{\fnm{Stefan} \sur{Klikovits} {\hypersetup{pdfborder={0 0 0}}\orcidlink{0000-0003-4212-7029}}}\email{stefan.klikovits@jku.at}

\author[3]{\fnm{Francis} \sur{Bordeleau} {\hypersetup{pdfborder={0 0 0}}\orcidlink{0000-0001-7727-3902}}}\email{francis.bordeleau@etsmtl.ca}

\author[1,4]{\fnm{Joachim} \sur{Denil} {\hypersetup{pdfborder={0 0 0}}\orcidlink{0000-0002-4926-6737}}}\email{joachim.denil@uantwerpen.be}

\author[5]{\fnm{{\O}ystein} \sur{Haugen} {\hypersetup{pdfborder={0 0 0}}\orcidlink{0000-0002-0567-769X}} }\email{oystein.haugen@hiof.no}

\affil*[1]{\orgdiv{Faculty of Applied engineering: Electronics and ICT}, \orgname{University of Antwerp}, \orgaddress{\street{Groenenborgerlaan 171}, \city{Antwerp}, \postcode{2020}, \state{Antwerp}, \country{Belgium}}}

\affil[2]{\orgdiv{Institute for Business Informatics - Software Engineering}, \orgname{Johannes Kepler University}, \orgaddress{\street{Altenberger Strasse 69}, \city{Linz}, \postcode{4040}, \country{Austria}}}

\affil[3]{\orgdiv{Department of Software Engineering and Information Technology}, \orgname{Ecole de techonologie supérieure (ETS)}, \city{Montreal}, \country{Canada}}

\affil[4]{\orgdiv{Flanders Make @UAntwerpen}, \orgname{University of Antwerp}, \orgaddress{\street{Groenenborgerlaan 171n}, \city{Antwerpen}, \state{Antwerp}, \postcode{2020},  \country{Belgium}}}

\affil[5]{\orgdiv{Department of Computer Science and Communication}, \orgname{{\O}stfold University College}, \orgaddress{\street{B R A veien 4}, \postcode{1757} \city{Halden}, \country{Norway}}}

\abstract{
Despite best efforts, various challenges remain in the creation and maintenance processes of \acp{dt}. One of those primary challenges is the constant, continuous and omnipresent evolution of systems, their user's needs and their environment, demanding the adaptation of the developed \ac{dt} systems. \Acp{dt} are developed for a specific purpose, which generally entails the monitoring, analysis, simulation or optimization of a specific aspect of an \emph{actual system}, referred to as the \AT. As such, when the twin system changes, that is either the AT itself changes, or the scope/purpose of a \ac{dt} is modified, the \acp{dt} usually evolve in close synchronicity with the AT. As \acp{dt} are software systems, the best practices or methodologies for software evolution can be leveraged.
This paper tackles the challenge of maintaining a (set of) DT(s) throughout the evolution of the user's requirements and priorities and tries to understand how this evolution takes place. In doing so, we provide two contributions: (i) we develop \emph{\DarTwin}, a visual notation form that enables reasoning on a twin system, its purposes, properties and implementation, and (ii) we introduce a set of architectural transformations that describe the evolution of \DT systems.
The development of these transformations is driven and illustrated by the evolution and transformations of a family home's \DT, whose purpose is expanded, changed and re-prioritized throughout its ongoing lifecycle. Additionally, we evaluate the transformations on a lab-scale gantry crane's \DT.

}

\maketitle

\glsresetall

\section{Introduction}\label{sec:intro}

The last two decades saw the adoption of \DTs in systems engineering in a wide range of application domains ~\citep{Glaessgen2012, kritzinger2018digital, Grieves2017, tao2022digital}. 

Many definitions of \DT can be found in the literature \citep{Glaessgen2012, kritzinger2018digital, dalibor2022cross}. In essence, a \DT is a digital representation of an actual system, referred to as the \footnote{While definitions of DT traditionally refer to Physical Twin, we prefer to use the term Actual Twin because DTs can be associated with different types of systems that are not exclusively physical, including processes, Cyber-Physical Systems (CPS) and socio-technical systems.}{\AT}, that is dynamically updated with AT data\footnote{The frequency of data update of the \DT depends on its purpose and goals.} \citep{wright2020, INCOSE2021}) and that can interact with and influence the \ac{at}. A DT is developed for a well-defined purpose, e.g. improve the energy efficiency of a building, and aims at achieving a set of goals related to its purpose. From an execution perspective, a DT uses different types of models \citep{eramo2021conceptualizing} to extract information and compute metrics from the collected data (current and historical) to provide a set of services \citep{eramo2021conceptualizing, tao2022digital} related to its purpose and goals. Examples of \DT services include the monitoring of specific properties, detection of system issues, data analytics to identify correlations between different system properties, simulation, and analysis of what-if scenarios.

Like any software artefact, \DTs are dynamic entities that evolve during their lifecycle to satisfy constantly evolving needs and requirements. 
The evolution of a \DT can be associated with three main types of modifications: 
\begin{enumerate}
    \item[\emph{i}] Modifications made to the \AT or its environment that need to be reflected in the \DT. For example, a \DT may need to be modified after a new sensor has been added to the \AT or a new external source of data has become available. A \DT may also need to be modified to reflect a structural modification made to the \AT.
    \item[\emph{ii}] Modifications made to the \DT to improve its precision with respect to one of its existing goals. For example, a \DT can be modified to take advantage of a new source of data (as a result of \emph{i}) or the introduction of a new type of model that can be used to improve its current thermal comfort. 
    \item[\emph{iii}] Modifications made to the purpose and/or goals of the \DT. For example, the scope of a \DT that was initially developed to improve thermal comfort in a building can be modified to include a new purpose or goal like the reduction of energy consumption and the improvement of air quality. This type of modification results in the creation of new \DT services.
\end{enumerate}

This last type of change (\emph{iii}) is related to the fact that \ATs are in general complex entities that have several aspects and that any \DT only addresses a specific subset of these aspects. While essentially all definitions of \DT refer to a digital representation/replica of an \AT, it would be more accurate to say that a \DT is a digital representation/replica of an \AT that only includes a subset of its aspects, for example, in the context of a smart building a \DT can be developed to improve thermal comfort while another \DTs can be developed to reduce energy consumption. In this article, we follow the view that a system can include several/many \DTs that each have a different purpose associated with specific aspects. We point out that our research is nonetheless compatible with alternative definitions, \eg where only one \DT exists, but it is decomposed into a set of components. In this context, during its lifecycle, the scope and purpose of a \DT may need to be progressively increased to include an increasing set of aspects of the \AT \citep{Arup2019, kamel2021digital}. 

Typically, when the evolution of a \DT is considered in the existing literature, it is concerned with the mirroring of the evolution of the \AT or its data \citep{lehner2021towards, chevallier2020reference}, \ie case (\emph{i}). However, doing so loses sight of the evolution of the purpose and scope of a \DT as it is given new, altered or improved purposes and goals, \ie case (\emph{iii}). For this case, we believe that \DTs must be architected and incrementally developed using a modular and scalable approach, just as it is done for systems that evolve over time. Hence, except for trivial cases, \DTs should be designed as encapsulated entities that can be composed, transformed, exchanged, and adapted to enable the incremental development of advanced functionality and sophisticated services.

While we appreciate the evolution of \DTs with respect to the three types of modifications, this paper focuses on the evolution of \DTs related to the modification made to their purpose and goals, i.e. (\emph{iii}). To support this type of evolution, we propose a systematic approach that is based on transforming the system architecture and synthesizing a set of tranformation templates. The approach aims at supporting the incremental development of a \DT (that addresses multiple aspects of a system) by composing \DTs that have been independently developed to address different purposes. Since the resulting set of \DTs is related to the same system, their integration needs to be carefully and systematically architected to avoid conflict and ensure optimal solutions.

As the goal of our study is to explore the evolution(s) of \DTs, case study research constitutes an appropriate research method to answer the proposed research questions~\citep{WOHLIN2021106514}. In this paper, we illustrate our approach using a smart home's heating system over the course of several requirements changes (\eg switch from thermal comfort prioritisation to considering energy consumption), and evaluate it with the help of a lab-scale gantry crane's \DT.

Specifically, our contributions are twofold: 
\begin{enumerate}
    \item Firstly, we develop the DarTwin notation that enables reasoning on a twin system, its purposes, properties and implementation.
    \item Secondly, we introduce a set of architectural transformations that we identified through analysis of our case study system
\end{enumerate}

The remainder of this paper is structured as follows: in \cref{sec:method} we discuss our underlying understanding of \DTs and introduce our views on the compositionality of \DTs. Next, in \cref{sec:experiment} we describe our notation for reasoning on the system's evolution, and the application on a smart home system through which we found the architectural transformations. Afterwards, in \cref{sec:evaluation}, we check if the architectural transformations are also suited for other systems, by applying them to a case study of a lab-scale gantry crane. Then, \cref{sec:discussion} positions our research and discusses some limitations. Following that, we discuss related work in \cref{sec:related}, and ongoing and future work in \cref{sec:pfw}, before concluding in \cref{sec:conclution}.

\section{Modular Aspects of DTs}\label{sec:method}

We have observed that most discussions seem to treat \DTs as relatively static artefacts.
Despite continuous (self-)adaptation using the \AT's data~\citep{lehner2021towards}, once created, the \DT's architecture typically remains conceptually unchanged.
As \ATs are usually part of a larger, continuously evolving system, it is evident that the corresponding \DTs should evolve too, \eg when new system parts are added, removed or restructured. Besides, continuous development and the addition of new features also requires the evolution of \DTs themselves.
Finally, we also note that the execution environment itself may change. We would call this process of managing the \DT evolution the \emph{\DT lifecycle management}.
Following this philosophy, we treat evolution as a continuous norm, where changes to any of the above aspects are omnipresent, occur on various granularity aspects and demand frequent adaptations. 
Hence, our goal should be to enhance current \DT practices and integrate a modern mindset that allows quick reaction and adjustment to such an evolution.

\subsection{DT Development and Evolution} 
We may think of \DT evolution in two ways. 
The first way deals with the evolution of an individual \DT, \eg based on updated data from the \AT (\eg~\citet{lehner2021towards}), the second way views the twin system more globally, in which we discuss the creation and interaction of \DTs and \ATs, or the interaction between \DTs (see \Cref{sec:experiment}).

Although this article focuses on the second way, we nonetheless briefly outline our understanding of an \AT-\DT system (\aka twin system) and its typical creation/design flow.

\begin{figure}[h]
\centering
\begin{tikzpicture}
\tikzset{
    every node/.append style={font=\sffamily},
    block/.append style={align=center,rectangle,draw=black,rounded corners=5pt},
    circlenum/.append style={pos=0.7,solid}
}

\node[rectangle,draw=black,fill=white,minimum width=8em,minimum height=3em] (system0) {Actual Twin (AT)};
\node[block,fill=white,minimum width=3em,minimum height=3em,above left=1 and .5 of system0.north] (dt1) {Digital Twin (DT)};
\node[block,fill=white,minimum width=3em,minimum height=3em, above right=1 and .5 of system0.north] (dt1b) {Digital Twin (DT)};

\begin{scope}[on behind layer]
    \node[sys,fit=(system0)(dt1)(dt1b),label={[below]north:Twin System 1 (TS)}] (system1) {};
\end{scope}

\draw[-latex,thick] (dt1.250) -- (system0.150)  node[circlenum,pos=.5,xshift=-1em] {2};
\draw[-latex,thick] (system0.120) -- (dt1.310) node[circlenum,pos=.6,xshift=1.5em] {1};

\draw[-latex,thick] (dt1.10) -- (dt1b.west |- dt1.10)  node[circlenum,pos=.5,below] {6};
\draw[-latex,thick] (dt1b.west |- dt1.-10) -- (dt1.-10);

\draw[-latex,thick] (dt1b.250) -- (system0.60)  node[circlenum,pos=.4,xshift=-1.5em] {5};
\draw[-latex,thick] (system0.30) -- (dt1b.310);

\umlactor[right=1.5 of system1] {User};
\draw[latex-latex,thick] (system1.east |- User) -- (User) node[circlenum,pos=0.5,yshift=1em] {3};

\node[block,fill=white,draw=gray,dashed,thick,minimum width=3em,minimum height=3em] (dt2) at ([xshift=-2cm]system1.west) {\textcolor{gray}{DT2}};

\draw[-latex,thick,dashed,gray] (dt2.25) -- (dt2.25 -| system1.west) node[circlenum,pos=.5,below] {4};
\draw[-latex,thick,dashed,gray] (dt2.-25 -| system1.west) -- (dt2.-25);

\begin{scope}[on background layer]
    \node[sys,dashed,fill=white,fit=(system1)(dt2),label={[below,gray]north:Twin System 2}] (system2) {};
\end{scope}

\end{tikzpicture}
\caption{Reference Architecture of a system-DT pair.}
\label{fig:reference-architecture2}
\end{figure}
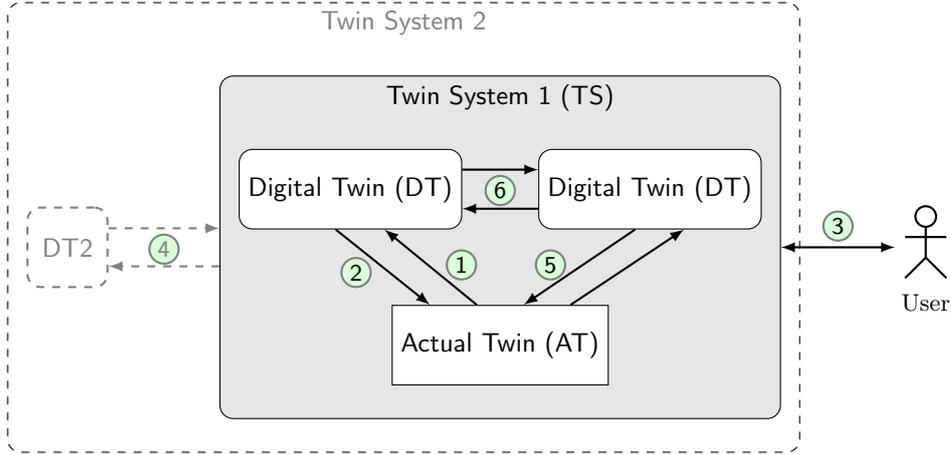

\Cref{fig:reference-architecture2} illustrates a typical twin system consisting of an \AT and a \DT, which interact in synergy, such that the \DT is updated with new data produced by the \AT \circlenum{1}. 
The feedback/control flow \circlenum{2} on the other hand is used to update the \AT, based on the implemented \DT services and the newly gained knowledge.

Next to the autonomous operation, twin systems typically enable user interaction \circlenum{3}. Note, that the figure indicates an interaction with the twin system as a unit, but in practice users may modify all parts of the system (\eg directly configure the \DT or alter the \AT's settings).

The \DT services may be clustered in a single \DT, or spread out over multiple \DTs. Each \DT only works with the subset of data and control flows it needs to provide its services \circlenum{5}, they may also rely on the services of another \DT to provide their own services \circlenum{6}.

Finally, we point out the hierarchical view that allows a twin system to be the \AT within another twin system. 
In the above figure, \textsf{Twin System 2} (dashed) contains \textsf{DT2} which serves as a \DT for \textsf{Twin System 1}, rather than the (inner) \textsf{\AT} or \textsf{\DT}. \circlenum{4} shows the communication from \textsf{DT2} and \textsf{Twin System 1} as an AT.
This conceptual distinction is important for avoiding conflicts of different \DT's purposes, as we will see further on in \Cref{sec:experiment}.

\subsection{Introduction to the Notation}

In order to better reason about the evolution of twin systems, including the above-mentioned purposes and \acp{poi}, we developed a notation called DarTwin, and the evolutionary scenarios are shown in framed diagrams with keyword \textbf{dartwin}.
Our notation aims to relate the \emph{why}, \emph{what} and \emph{how} of a certain evolution. Specifically:
\begin{enumerate}
    \item[\emph{Why?}] What are the goals of the twin system, and how are they evolving?
    \item[\emph{What?}] What \acfp{poi} of the twin system are we working on? 
    \item[\emph{How?}] What is the architecture of the twin system, and how does it relate to the goals and properties of interest?
\end{enumerate}

\begin{figure}[htb]
\centering
\resizebox{\linewidth}{!}{
\begin{tikzpicture}[dartwinFramed={\emph{example system}}{}]
\tikzset{
fontsize/.append style={font=\fontsize{8}{8}\selectfont\sffamily}
}
    \node[goal,fontsize,inner sep=0.5em,minimum height=2.5em,inner ysep=.5em] (goal) at (-1,0) {\textbf{Goal}\\PoIs\\ Constraints};

    \node[process,fontsize, below=1.75 of goal,minimum width=5em,yshift=1em] (dt) {Digital Twin};
    \pic (schema-arbiter) at ($(dt.south)+(3,.0)$){schema arbiter dt={Arbiter/}};

    \node[port,label={[portlabel]above left:{Ports}}] (dt-port-in) at (dt.west) {$\rightarrow$};
    \node[port,label={[portlabel]left:{}}] (dt-port-out) at (dt.east) {$\rightarrow$};

    \begin{scope}[on behind layer]
       \node[sys,fill=gray!20,label={[below]north:{\scriptsize \textbf{twin system} \emph{systemname}}},minimum width=6.5cm,minimum height=3cm,below right=1.75 and -1.25 of goal.west] (system) {};
    \end{scope}
    \node[port,label={[portlabel]left:{}}] (sys-port) at (system.west |- dt-port-in) {$\rightarrow$};
    \node[port,label={[portlabel]below:{}}] (sys-port-out) at (system.south -| schema-arbiter-out) {$\downarrow$};

    \draw[flow] (sys-port) -- (dt-port-in) node[below,midway,arrowlabel] {};
    \draw[flow] (schema-arbiter-out) -- (sys-port-out) node[below,midway,arrowlabel] {};

    \draw[flow] (dt-port-out) -- (schema-arbiter-in1.north) node[above,midway,arrowlabel] {Flow};
    \draw[relation] (dt) -- (goal);

\pic {frameSeparation={goal}};  
\end{tikzpicture}
\begin{tikzpicture}  
\tikzset{
fontsize/.append style={font=\fontsize{8}{8}\selectfont\sffamily}
}
    \draw[specialization] (0,0) -- ++(1,0);
        \node[arrowlabel,fontsize,align=left] (label1) at (3,0.25) {Generalisation};    
    \draw[goalrelation] (0,-.5) -- ++(1,0);
        \node[arrowlabel,fontsize,align=left] (label2) at (3,-.25) {Goal Relation};
    \draw[conflict] (0,-1) -- ++(1.25,0);
        \node[arrowlabel,fontsize] (label3) at (3,-.75) {Conflict beteen Goals};

\begin{scope}[yshift=-3.5cm]  
    \draw[relation] (0,0) -- ++(1,0);
        \node[arrowlabel,fontsize,align=left] (label4) at (3,.25) {Relation between DT and Goal};
    \draw[flow] (0,-.5) -- ++(1,0);
        \node[arrowlabel,fontsize,align=left] (label5) at (3,-.25) {Flow between Ports};

    \node[] at (0,-2) {};  
\end{scope}
\end{tikzpicture}
}

\caption{Legend containing the building blocks used in the \DarTwin notation.}
\label{fig:dependencies-legend}
\end{figure}
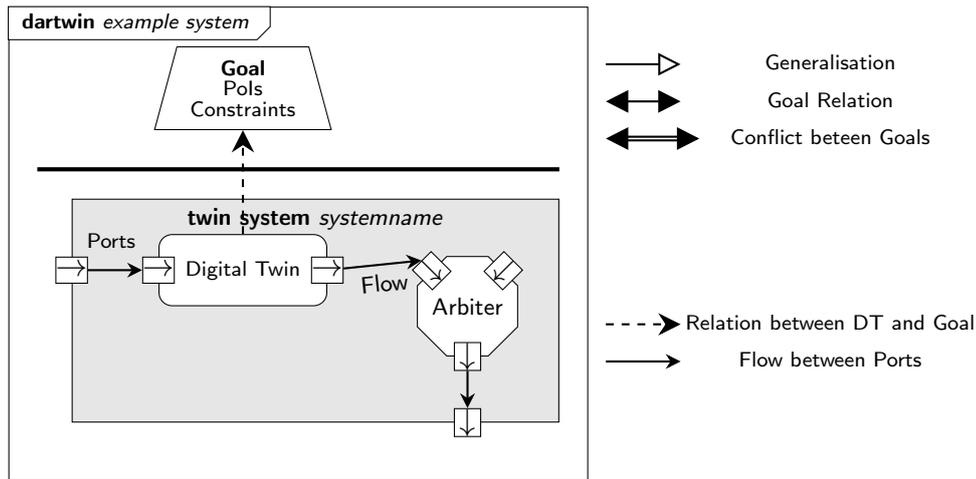

The legend containing the elements of our notation is shown in \cref{fig:dependencies-legend}. The two main elements are the trapezoidal-shaped goal and the roundtangle shaped DT. The diagram is split in two sections by a horizontal line, as shown in the legend as well. Above that horizontal separator, the trapezoids will form the whole of the goals of the twin system, whilst below, the roundtangles will form a communication diagram of the twin system architecture.

The goal trapezoid contains three elements: a name/title describing the goal, the \acfp{poi} related to that goal, and the constraints to which must be adhered to fulfill the goal. 

The \Acp{poi} are characteristics of the AT or its environment that designers consider to evaluate whether a specific goal is reached or not. For example, in the context of building improvement, we could be interested in the room temperature, level of pollution in the air or the energy consumption. The \acp{poi} allow us to abstract aspects of the AT and reason about them in the goals. The \Ac{poi} is typically retrieved as payloads of the messages that pass through the ports of the DT roundtangles. This represents the four-variable model introduced by Parnas and Madey~\citep{PARNAS199541}.

The constraints describe how the \Ac{poi} must behave for the goal to be fulfilled.

During an evolution, goals appear in three different flavours: 
\begin{enumerate}[i)]
    \item orthogonal, where the new goal has no connection to the current goals of the twin system;
    \item extending, such that the new goal extends the current goal(s) of the system, depicted using the ``generalisation relation'' arrow; 
    \item positive relation or opposing/conflicting relation, the new goal positively affects or opposes/conflicts with one or more of the current goals of the system, shown using either a ``goal relation'' or ``conflict arrow'' between the goals. 
\end{enumerate}

This then leads us to the roundtangle shaped DT. It contains a name of the DT it represents, as well as a set of communication ports. Arrows indicate the flow of data between the ports. All linked together, they form a communication diagram of the twin system architecture, similar to composite structures in UML \citep{OMG-UML}. Dashed relations connect \DTs of the architecture with their corresponding goals of the conceptual upper part. These dashed arrows are the only element that crosses the horizontal separator.

\subsection{Architectural Transformations}

We will use our running example in the next section to explain our approach and to depict the evolutions. We apply an orange colour (visually impaired-friendly) to depict the changes/additions. For each evolution, we try and summarize our motivation by an architectural transformation that generalizes the evolution. The transformation applies the same notation as the example itself, in a simpler and more generic diagram.

In general, our approach is that of separation of concerns. The concerns are first defined by the goals, and then the implementation will typically see whether the new concern can be represented by a separate \DT. Alternatively, we do changes to the architecture, and modifications of existing \DTs.

\section{Twin Systems and their structured evolution}\label{sec:experiment}
In the following sections, we introduce our running example and  describe its evolution over the course of several purpose switches using the graphical notation.

\subsection{Running Example: A Smart Home System} 

\begin{figure}
    \centering
    \includegraphics[width=\textwidth]{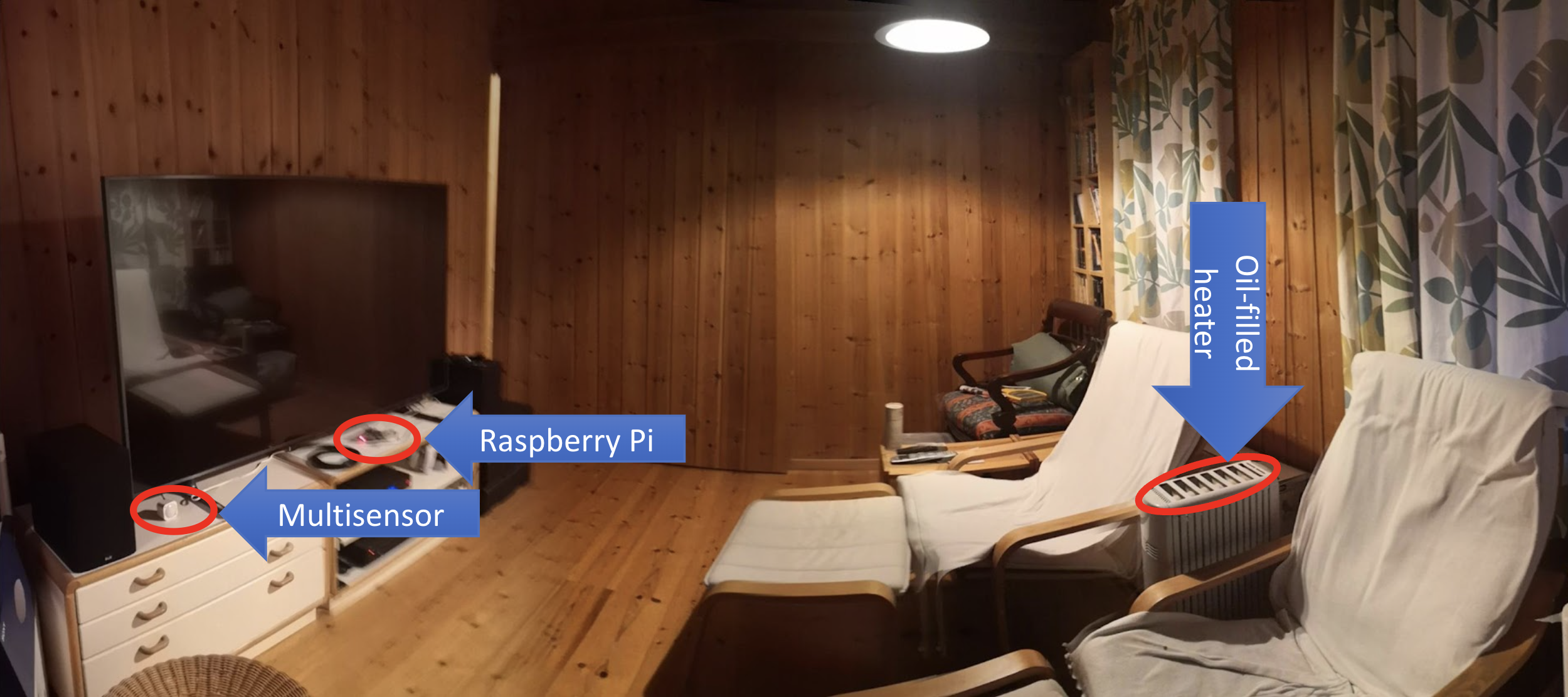}
    \caption{The case study system is a room in a Norwegian home, equipped with multi-sensor, Raspberry Pi controller and an oil-filled heater.}
    \label{fig:room-img}
\end{figure}

We introduce our approach using a running example that is based on an actually existing, single-family home in Norway. Despite focusing on one individual room to explain our concepts and the evolution of the system, in fact, several rooms in the house are controlled by the developed solution. The house itself was originally built in 1965 and maintained with several independent climate-controlling systems such as an invertible heat pump and mechanical ventilation. 
As these installations were made over the course of several decades, unfortunately, these systems are independent, that is, without a common control. The focus of our running example is on one room, that is equipped with one electric oil-filled heater (see \Cref{fig:room-img}). A multisensor capable of sensing temperature, luminance, humidity and motion is installed to provide sensory input to a Raspberry Pi that serves as an independent control unit of the room by actuating a power switch of the heater.

\subsubsection{The Basic System: Thermal Comfort} 

Due to traditionally low energy prices in Norway, the original and main purpose of controlling the thermal comfort was (and mostly remains) to keep the room temperature pleasant for occupants. Furthermore, given the climatic circumstances (long and fairly cold winters) the main functionality requires controlling the room heating (without active cooling).
Thus, the first logic for controlling the room was a simple thermostat where we defined the comfort temperature and the allowed deviation from that comfort temperature.

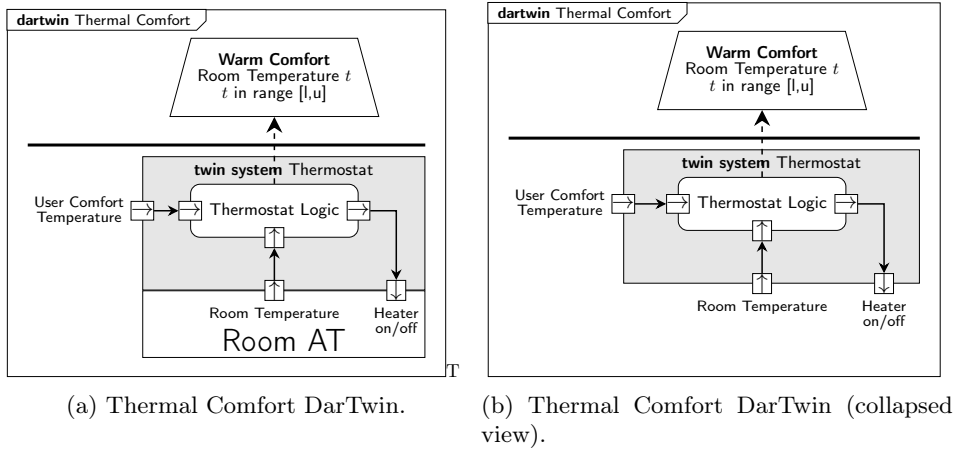
\begin{figure}
\begin{subfigure}[t]{.475\textwidth}
\centering
\resizebox{\linewidth}{!}{
\begin{tikzpicture}[dartwinFramed={Thermal Comfort}{}]
\pic {comfort goal};

\begin{scope}[yshift=-2.5cm]  
    \pic (comfort-dt) {comfort dt};
    \begin{scope}[on behind layer]
       \node[sys,fill=gray!20,label={[below]north:{\scriptsize \textbf{twin system} Thermostat}},minimum width=5.25cm,minimum height=2.5cm,xshift=.5em] (system) at (0,-.5) {};
    \end{scope}
    
    \pic (comfort-wrap) {comfort wrap ports hooked={system}};
    \draw[flow] (comfort-wrap-comfort-temp) -- (comfort-dt-comfort-temp); 

    \begin{scope}[on behind layer]
    \node[actualtwin,minimum width=5.25cm,minimum height=1.25cm,below=0 of system,label={[above]south:{\Large Room AT}}] (room AT) {};
    \end{scope}
\end{scope}

\draw[relation] (comfort-dt-dt) -- (comfort goal);

\pic {frameSeparation={comfort goal}};  
\end{tikzpicture}T
}
\caption{Thermal Comfort DarTwin.}
\label{fig:warm_comfort}
\end{subfigure} %
\begin{subfigure}[t]{.475\textwidth}
\centering
\resizebox{\linewidth}{!}{
\begin{tikzpicture}[dartwinFramed={Thermal Comfort}{}]
\pic {comfort goal};

\begin{scope}[yshift=-2.5cm]  
    \pic (comfort-dt) {comfort dt};
    \begin{scope}[on behind layer]
       \node[sys,fill=gray!20,label={[below]north:{\scriptsize \textbf{twin system} Thermostat}},minimum width=5.5cm,minimum height=2.5cm,xshift=.5em] (system) at (0,-.5) {};
    \end{scope}
    
    \pic (comfort-wrap) {comfort wrap ports hooked={system}};
    \draw[flow] (comfort-wrap-comfort-temp) -- (comfort-dt-comfort-temp); 

    \node[below=1.15cm of system] (dummy) {};
\end{scope}

\draw[relation] (comfort-dt-dt) -- (comfort goal);

\pic {frameSeparation={comfort goal}};  
\end{tikzpicture}
}
\caption{Thermal Comfort DarTwin (collapsed view).}
\label{fig:warm_comfort_collapsed}
\end{subfigure}
\caption{Two views of the Thermal Comfort DarTwin.}
\label{fig:warm_comfort_two_views}
\end{figure}

\Cref{fig:warm_comfort} shows the \textsf{Thermal Comfort} DarTwin with the \textsf{Warm Comfort} purpose with the temperature \ac{poi}, above the horizontal separation bar. Here the goal is to provide \textsf{Warm Comfort} for occupants and the way we operationalize it is to control the \textsf{Room Temperature} as indicated in the goal with a simple constraint giving a range for the room temperature.

When it comes to the actual implementation, we may view the room altogether as a system with temperature control. However, another view can designate the Raspberry Pi program with inputs from the multisensor and outputs to the \textsf{Heater}'s switch actuator, as a virtual representation (or a \DT) of the physical room with its heater. We therefore denote the whole room, including the controller, as \textsf{Thermostat} twin system, consisting of one physical system (the \textsf{Room} \AT) and one virtual system (the \textsf{Thermostat Logic} \DT).

The implementation of this initial room is depicted in \Cref{fig:warm_comfort}, which shows the \AT being the \textsf{Room} \AT with \textsf{Temperature Sensor} for observing temperature and an oil-filled electric \textsf{Heater} to be controlled by the \textsf{Thermostat Logic} \DT's simple on or off actions. A user can set the desired comfort temperature through the \textsf{User Comfort Temperature} input.
The dashed arrow crossing the bar shows the dependency between the goal specification and its \DT implementation. We depicted the \textsf{Room} \AT as a rectangle, but realize that it is normally not necessary to indicate the \AT with more than its communication ports. Indeed, the \AT can be collapsed as shown in \cref{fig:warm_comfort_collapsed}. In \Cref{fig:warm_comfort} we have the \textsf{Room Temperature} sensor and the \textsf{Heater} actuator, and we decide to place those ports on the bottom edge of the \textsf{Thermal Comfort} twin system. In the remainder of the paper, we default to the collapsed view unless otherwise stated.

\subsubsection{The initial transformation template}
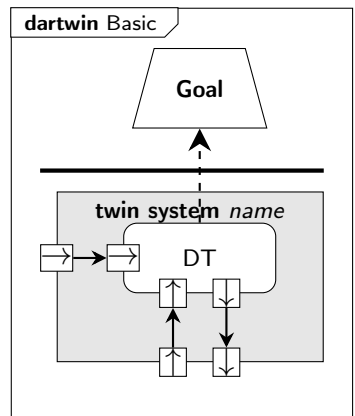
\begin{figure}
\centering
\begin{tikzpicture}[dartwinFramed={Basic}{}]
\pic (goal0) {schema goal={Goal/}};

\begin{scope}[yshift=-2cm]  
    \pic (schema-dt) at (0, -.25){schema dt={DT/}};
    \begin{scope}[on behind layer]
       \node[system,fill=gray!20,label={[below]north:{\scriptsize \textbf{twin system} \textit{name}}},minimum width=3.5cm,minimum height=2.25cm] (system) at (-.125cm,-.5) {};
    \end{scope}
    
    \pic (system-inout) {schema aligned inout={system/schema-dt/}};

    \draw[flow] (system-inout-in.north) -- (schema-dt-in.south);
    \draw[flow] (schema-dt-out.south) -- (system-inout-out.north);
\end{scope}

\node[port,label={below:{\scriptsize }}] (left-in) at (schema-dt-dt.west) {$\rightarrow$};
\node[port,label={below:{\scriptsize }}] (system-left-in) at (system.west |- schema-dt-dt.west) {$\rightarrow$};

\draw[flow] (system-left-in) -- (left-in);

\draw[relation] (schema-dt-dt) -- (goal0-goal);

\pic {frameSeparation={goal}};  
\end{tikzpicture}
\caption{Basic generic DarTwin.}
\label{fig:schema-new-goal}
\end{figure}

The most basic template is given in \Cref{fig:schema-new-goal}. It merely shows a typical starting design with one specific goal implemented through one \DT. Typically, there is communication with the \AT through ports representing sensors and actuators. The sensor ports carry monitoring data while the actuator ports carry control data. Most systems will have input from the users, and may be also output to the user (not shown here).

\subsection{Evolution of the Twin System}

After the initial configuration of the heating's digital operation, over the years the system underwent several significant modifications. 
Step by step, new system purposes were introduced, thereby requiring iterative adaptation and extension of the existing \DT and system.

\subsubsection{Evolution 1a: Green Comfort with hierarchical transformation} 

Despite the comparatively low-cost energy, even in Norway, a wave of energy-awareness led to the ambition to decrease the energy use in homes to contribute in general to a better environment.
Thus, a new goal was introduced into the system, namely that the room should be more energy-efficient. 
By considering the existing \textsf{Thermostat} twin system as the \AT, the aim was to manipulate the parameters of the thermostat to lower energy consumption without sacrificing the occupants' thermal comfort. 
Hence, it was decided to lower the comfort temperature when the room was not used (\ie nobody is in the room), or when the room was used for sleeping. 

\begin{figure}
\centering
\resizebox{\textwidth}{!}{
\begin{tikzpicture}[dartwinFramed={Green Comfort}{Thermal Comfort}]

\pic {comfort goal};
\pic [left=3.2 of comfort goal] {energy goals={highlight}};
\draw[goalrelation,highlight] (comfort goal) -- (lowerenergy goal) node[midway,above,] {\scriptsize Lower e $\Leftrightarrow$ Lower t};

\begin{scope}[yshift=-5.5cm,xshift=-2cm]  
    \begin{scope}[on behind layer]
       \node[system,fill=gray!5,label={[below]north:{\textbf{twin system} Energy Saving Thermostat}},minimum width=11cm,minimum height=2.75cm] (energysave-system) at (-2,-.25) {};
    \end{scope}
    
    \pic (energysave-dt) at (-5, 0) {energysave dt={highlight}};
    
    \node[port,highlight,label={[portlabel]below:{Presence}}] (-presence) at (energysave-system.south -| energysave-dt-presence) {$\uparrow$}; 
    \draw[flow,highlight] (-presence) -- (energysave-dt-presence);
    
    \node[port,label={[portlabel]left:{User Comfort\\Temperature}}] (inner-system-comfort-temp) at ($(energysave-system.south)$) {$\downarrow$};

    \begin{scope}[on behind layer]
        \node[system,fill=gray!20,label={[below]north:{\textbf{twin system} Thermostat as AT}},minimum width=6cm,minimum height=3cm,below left=0 and 0 of energysave-system.south east] (system) {};
        \pic (comfort-dt) at (system) {comfort dt};
        \pic (comfort-wrap) {temp heat ports hooked={system}};
    \end{scope}
    
    \draw[flow] (inner-system-comfort-temp) |- (comfort-dt-comfort-temp); 
\end{scope}

\node[port,highlight,label={[portlabel]left:{User Comfort\\Temperature}}] (-comfort-temp) at (energysave-system.west |- energysave-dt-comfort-temp-in) {$\rightarrow$};
\draw[flow,highlight] (-comfort-temp) -- (energysave-dt-comfort-temp-in);
\draw[flow,highlight] (energysave-dt-comfort-temp-out) -| (inner-system-comfort-temp);

\draw[relation] (comfort-dt-dt) -- (comfort goal);
\draw[relation,highlight] (energysave-dt-dt) -- (whenabsent goal);

\pic {frameSeparation={whenabsent goal}};  
\end{tikzpicture}
}
\caption{Green Comfort DarTwin introduces the additional goal \textsf{Lower Energy} which is satisfied by the \textsf{Energy Saving} \DT.}
\label{fig:green_comfort}
\end{figure}
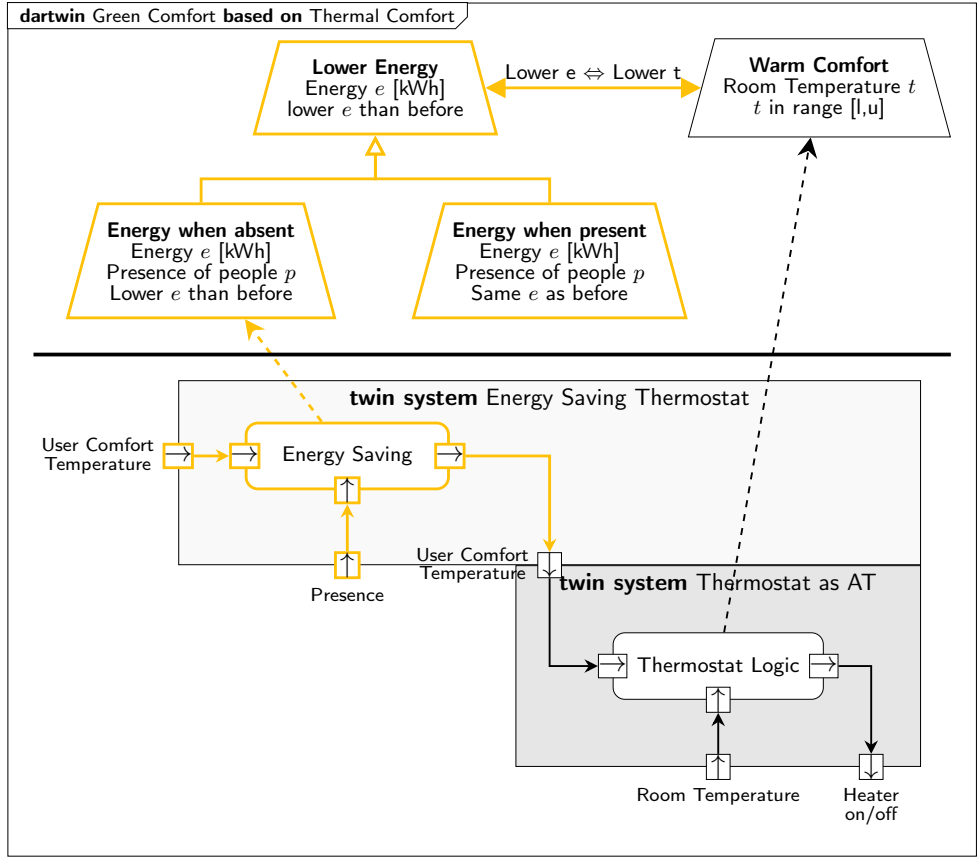

We have depicted our desire and energy-saving reasoning, which we call the \textsf{Green Comfort} evolution, in \Cref{fig:green_comfort} by showing the incremental changes in orange colour. We specialize the \textsf{Lower Energy} goal into two separate disjunct cases, namely when people are absent(or sleeping at night) and when they are present. When people are absent, we may lower the energy usage without affecting the thermal comfort. When there are people present, we keep the previous notion of thermal comfort.

Therefore, we designed a controlling process that adjusted the user's comfort temperature based on predictions of the room's use. 
A luminosity sensor (from the multisensor) is used to indicate occupancy.
When the room is lit, we assume there are people present or intend to become present and therefore the thermal comfort level should be normal. On the other hand, when there is no light, the thermostat's temperature is lowered. 
Furthermore, a distinction between day and night keeps the temperature slightly higher during the day, so the time to heat to the user's comfort temperature is shortened once the light is turned on.

We now have two different perspectives on the \textsf{Green Comfort} evolution. Either we see the \textsf{Thermostat} twin system as a finished system that we will see as the \AT, or we see the original \textsf{Thermostat} twin system as a system that we augment by one more \DT. We show both alternatives of this new \textsf{Energy Saving Thermostat} twin system.

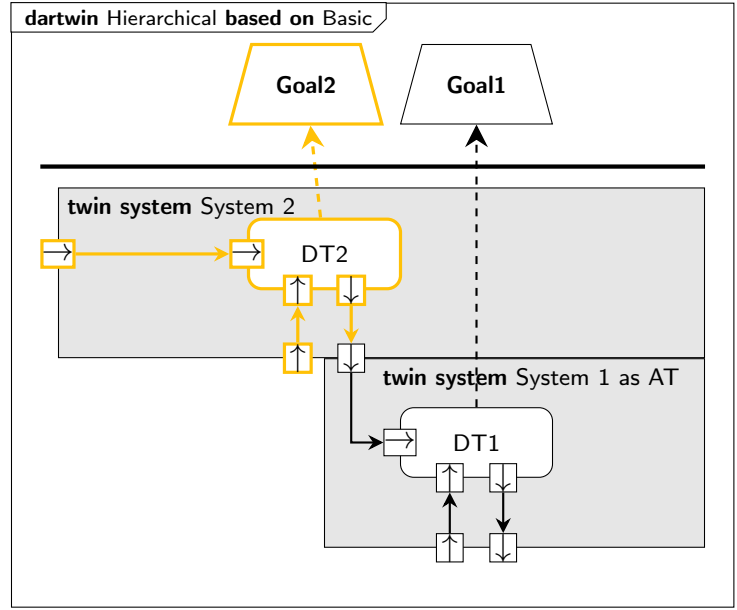
\begin{figure}
\centering
\begin{tikzpicture}[dartwinFramed={Hierarchical}{Basic}]

\pic (goal1) {schema goal={Goal1/}};
\pic [left=.5 of goal1-goal] (goal2) {schema goal={Goal2/highlight}};

\begin{scope}[yshift=-2cm]  
    \begin{scope}[on behind layer]
       \node[system,fill=gray!20,label={[below right]north west:{\scriptsize \textbf{twin system} System 2}},minimum width=8.5cm,minimum height=2.25cm] (system2) at (-1.25cm,-.5) {};
    \end{scope}
    
    \pic (schema-dt2) at (-2, -.25){schema dt={DT2/highlight}};
    \pic (system2-inout) {schema aligned inout-highlight={system2/schema-dt2/highlight}};
    \draw[flow,highlight] (system2-inout-in.north) -- (schema-dt2-in.south);
    \draw[flow,highlight] (schema-dt2-out.south) -- (system2-inout-out.north);
    
    \node[port,highlight,label={below:{\scriptsize }}] (left-in) at (schema-dt2-dt.west) {$\rightarrow$};
    \node[port,highlight,label={below:{\scriptsize }}] (system-left-in) at (system2.west |- schema-dt2-dt.west) {$\rightarrow$};
    \draw[flow,highlight] (system-left-in) -- (left-in);
    
    \begin{scope}[on behind layer]
       \node[system,fill=gray!20,label={[below right]north west:{\scriptsize \qquad\textbf{twin system} System 1 as AT}},minimum width=5cm,minimum height=2.5cm,below left=0 and 0 of system2.south east] (system) {};
    \end{scope}
    \pic (schema-dt1) at (0, -2.75){schema dt={DT1/}};
    \pic (system-inout) {schema aligned inout={system/schema-dt/}};
    \draw[flow] (system-inout-in.north) -- (schema-dt1-in.south);
    \draw[flow] (schema-dt1-out.south) -- (system-inout-out.north);

    \node[port,label={below:{\scriptsize }}] (dt1-left-in) at (schema-dt1-dt.west) {$\rightarrow$};

    \draw[flow] (system2-inout-out.south) |- (dt1-left-in);

\end{scope}

\draw[relation] (schema-dt1-dt) -- (goal1-goal);
\draw[relation,highlight] (schema-dt2-dt) -- (goal2-goal);

\pic {frameSeparation={goal}};  
\end{tikzpicture}
\caption{Hierarchical transformation describing a nested twin system.}
\label{fig:schema-hierarchical}
\end{figure}

There may be reasons to consider the simple thermostat system as a fixed system. It may have been deployed in a way that we have only very limited ways to manipulate it. We assume our only communication with the original \textsf{Thermostat} is through giving the user comfort temperature. Thus, our process serves as a \DT relative to our \textsf{Thermostat} twin system, effectively establishing a hierarchical, nested twin system as depicted in \Cref{fig:green_comfort}. We introduce a presence sensor, which abstracts over the combination of luminance and time of day. We see that the new \textsf{Energy Saving} \DT is connected to the existing \textsf{Thermostat} twin system. This addition corresponds to what we call a hierarchical transformation pattern, as shown generalized in \Cref{fig:schema-hierarchical}.

The hierarchical pattern represents an evolution where the original twin system is treated  as a black box. Treating a twin system as a black box means that, from the new twin system's vantage point, we consider it as the \AT. This means that the user inputs and outputs of the original twin system now become the controlling and monitoring data of the new twin system.

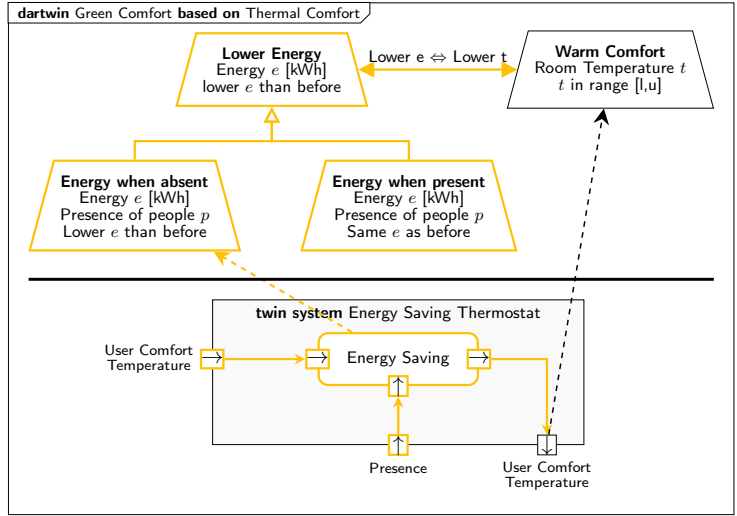
\begin{figure}
\centering
\resizebox{.75\textwidth}{!}{
\begin{tikzpicture}[dartwinFramed={Green Comfort}{Thermal Comfort}]

\pic {comfort goal};
\pic [left=3 of comfort goal] {energy goals={highlight}};
\draw[goalrelation,highlight] (comfort goal) -- (lowerenergy goal) node[midway,above,] {\scriptsize Lower e $\Leftrightarrow$ Lower t};

\begin{scope}[yshift=-5.5cm]  

    \begin{scope}[on behind layer]
       \node[system,fill=gray!5,label={[below]north:{\scriptsize \textbf{twin system} Energy Saving Thermostat}},minimum width=7cm,minimum height=2.75cm] (energysave-system) at (-4,-.25) {};
    \end{scope}
    
    \pic (energysave-dt) at (-4, 0) {energysave dt={highlight}};
    
    \node[port,highlight,label={[portlabel]below:{Presence}}] (-presence) at (energysave-system.south -| energysave-dt-presence) {$\uparrow$}; 
    \draw[flow,highlight] (-presence) -- (energysave-dt-presence);

    \node[port,label={[portlabel]below:{User Comfort\\Temperature}},xshift=-2em] (inner-system-comfort-temp) at (energysave-system.south east) {$\downarrow$};
\end{scope}

\node[port,highlight,label={[portlabel]left:{User Comfort\\Temperature}}] (-comfort-temp) at (energysave-system.west |- energysave-dt-comfort-temp-in) {$\rightarrow$};
\draw[flow,highlight] (-comfort-temp) -- (energysave-dt-comfort-temp-in);
\draw[flow,highlight] (energysave-dt-comfort-temp-out) -| (inner-system-comfort-temp);

\draw[relation] (inner-system-comfort-temp) -- (comfort goal);
\draw[relation,highlight] (energysave-dt-dt) -- (whenabsent goal);

\pic {frameSeparation={whenabsent goal}};  
\end{tikzpicture}
}
\caption{Green Comfort DarTwin with the \AT bottom view.}
\label{fig:green_comfort_at_bottom}
\end{figure}

Since there are no modifications to the original \textsf{Thermostat} twin system in this evolution, we do not need to show it in full detail. We can just represent it by one port for \textsf{User Comfort Temperature} as depicted in \Cref{fig:green_comfort_at_bottom}. This may help focus on the elements we now design and evolve.

\subsubsection{Evolution 1b: Green Comfort with augmented transformation} 
\label{sec:greentwinflat}

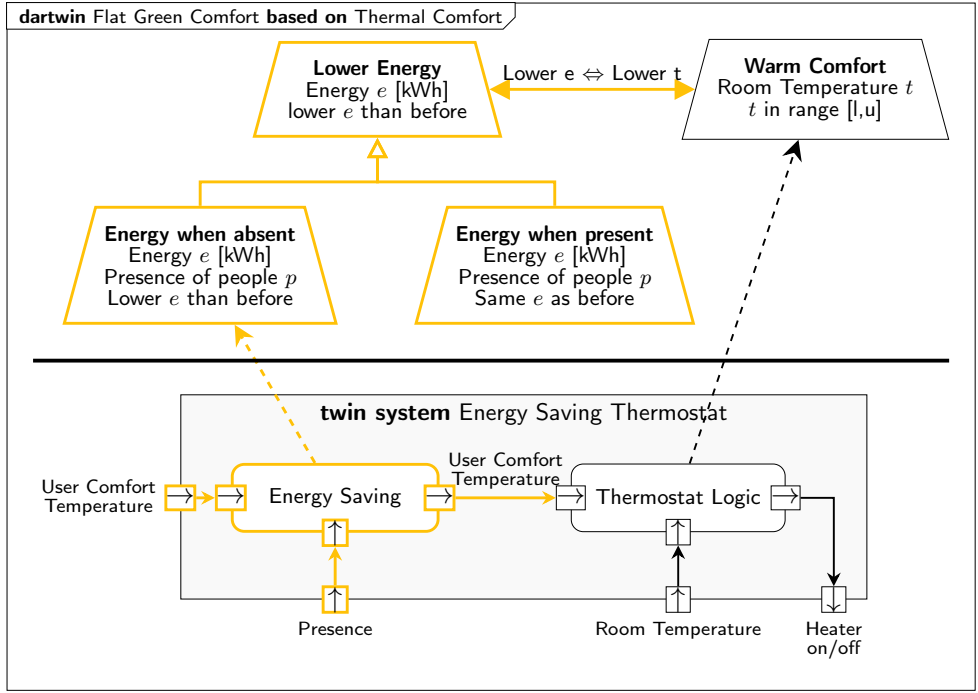
\begin{figure}[htb]
\centering
\resizebox{\textwidth}{!}{
\begin{tikzpicture}[dartwinFramed={Flat Green Comfort}{Thermal Comfort}]
\pic {comfort goal};
\pic [left=3 of comfort goal] {energy goals={highlight}};
\draw[goalrelation,highlight] (comfort goal) -- (lowerenergy goal) node[midway,above,] {\scriptsize Lower e $\Leftrightarrow$ Lower t};


\begin{scope}[yshift=-6cm,xshift=-2cm]
    \begin{scope}[on behind layer]
       \node[sys,fill=gray!5,label={[below]north:{\textbf{twin system} Energy Saving Thermostat}},minimum width=10cm,minimum height=3cm] (energysave-system) at (-2.25,-.25) {};
    \end{scope}

    \pic (energysave-dt) at (-5, 0) {energysave dt={highlight}};
    
    \node[port,highlight,label={[portlabel]below:{Presence}}] (outer-presence) at (energysave-system.south -| energysave-dt-presence) {$\uparrow$};
    \draw[flow,highlight] (outer-presence) -- (energysave-dt-presence);

    \pic (comfort-dt) {comfort dt};
    \pic (comfort-wrap-outer) {comfort wrap ports hooked-highlight={energysave-system}};

    \draw[flow,highlight] (energysave-dt-comfort-temp-out) -- node[portlabel,above,midway]{User Comfort\\Temperature} (comfort-dt-comfort-temp); 
\end{scope}

\draw[flow,highlight] (comfort-wrap-outer-comfort-temp) -- (energysave-dt-comfort-temp-in);

\draw[relation] (comfort-dt-dt) -- (comfort goal);
\draw[relation,highlight] (energysave-dt-dt) -- (whenabsent goal);

\pic {frameSeparation={whenabsent goal}};  
\end{tikzpicture}
}
\caption{Flat Green Comfort DarTwin applying the augmented transformation.}
\label{fig:green_comfort_refact_flattened}
\end{figure}

The hierarchical \textsf{Energy Saving Thermostat} nested twin system, could have been deployed with the two \DTs on different physical hardware, and in principle, the \textsf{Thermostat} twin system could have been available to the programmer of the \textsf{Energy Saving} \DT only through a limited interface. In our real twin system, however, the border of the inner \textsf{Thermostat} twin system is defined by software running on the Raspberry Pi. Therefore, for a systems engineer it is natural to consider whether the new \DT should just augment the twin system as a parallel \DT, as depicted in \Cref{fig:green_comfort_refact_flattened}. 

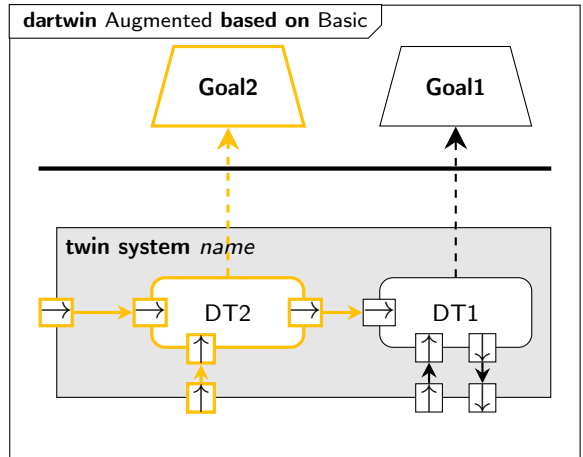
\begin{figure}
\centering
\begin{tikzpicture}[dartwinFramed={Augmented}{Basic}]

\pic (goal1) {schema goal={Goal1/}};
\pic [left=1.25 of goal1-goal] (goal2) {schema goal={Goal2/highlight}};

\begin{scope}[yshift=-3cm]  
    \begin{scope}[on behind layer]
       \node[system,fill=gray!20,label={[below right]north west:{\scriptsize \textbf{twin system} \textit{name}}},minimum width=6.5cm,minimum height=2.25cm] (system2) at (-2cm,0) {};
    \end{scope}

    \pic (schema-dt2) at (-3, 0){schema dt only={DT2/highlight}};
    \node[port,highlight,label={below:{\scriptsize }}] (dt2-left-in) at (schema-dt2-dt.west) {$\rightarrow$};
    \node[port,highlight,label={below:{\scriptsize }}] (system-left-in) at (system2.west |- schema-dt2-dt.west) {$\rightarrow$};
    \draw[flow,highlight] (system-left-in) -- (dt2-left-in);

    \node[port,highlight,label={below:{\scriptsize }},xshift=-1em] (dt2-south-in) at (schema-dt2-dt.south) {$\uparrow$};
    \node[port,highlight,label={below:{\scriptsize }},xshift=-1em] (system2-south-in) at (system2.south -| schema-dt2-dt.south) {$\uparrow$};
    \draw[flow,highlight] (system2-south-in) -- (dt2-south-in);
    
    \node[port,highlight,label={below:{\scriptsize }}] (system-south-in) at (system2.south -| dt2-south-in) {$\uparrow$};
    \node[port,highlight,label={below:{\scriptsize }}] (dt2-right-out) at (schema-dt2-dt.east) {$\rightarrow$};

    \pic (schema-dt1) at (0, 0){schema dt={DT1/}};
    \pic (system-inout) {schema aligned inout={system2/schema-dt/}};

    \draw[flow] (system-inout-in.north) -- (schema-dt1-in.south);
    \draw[flow] (schema-dt1-out.south) -- (system-inout-out.north);
    
    \node[port,label={below:{\scriptsize }}] (dt1-left-in) at (schema-dt1-dt.west) {$\rightarrow$};
    \draw[flow,highlight] (dt2-right-out) -- (dt1-left-in);

\end{scope}

\draw[relation] (schema-dt1-dt) -- (goal1-goal);
\draw[relation,highlight] (schema-dt2-dt) -- (goal2-goal);

\pic {frameSeparation={goal}};  
\end{tikzpicture}
\caption{Augmented transformation.}
\label{fig:schema-flat}
\end{figure}

The connection between the \textsf{Energy Saving} and the \textsf{Thermostat Logic} \DTs is now internal to the \textsf{Energy Saving Thermostat} twin system and can be adjusted at will without affecting the interface to the \textsf{Room} \AT. We have called this transformation \emph{Augmented} as depicted in \Cref{fig:schema-flat}. The augmented transformation may rely on plain communication between the constituent \DTs. This means that the inputs and outputs of the constituents may interconnect, while the monitoring and controlling data of the \DTs will be kept separate for each \DT. 

The \textsf{Flat Green Comfort} evolution applying the augmented transformation shown in \Cref{fig:green_comfort_refact_flattened} corresponds with our real chosen implementation and we shall continue from that variant in the next evolution. 

\subsubsection{Evolution 2a: Freeze Protection with orthogonal transformation} 
\label{sec:freezeprotect}

Nowadays, winters are often warmer than before, but Norway is still cold sometimes. A risk analysis of our programmed temperature control system revealed the need to cope with extreme cases. Every year water pipes in Norway freeze and subsequently break with very unpleasant consequences for the owners. In particular this often happens in remote or vacation cabins where the people are not always present to realize the catastrophe. This is why we decided to introduce an independent \textsf{Freeze Protection} \DT process. Often also called ``watchdogs'', such independently functioning monitoring processes should be as independent of the core system as possible. 

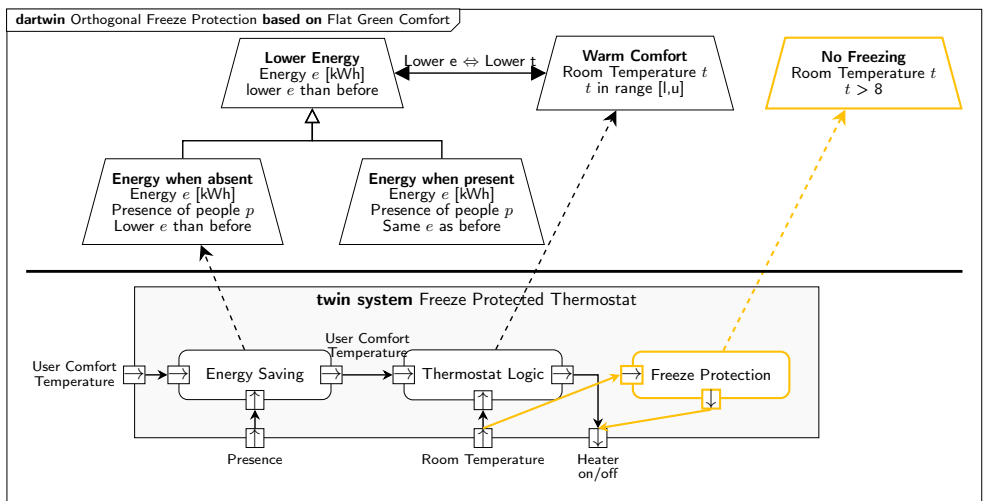
\begin{figure}[htb]
\centering
\resizebox{\textwidth}{!}{
\begin{tikzpicture}[dartwinFramed={Orthogonal Freeze Protection}{Flat Green Comfort}]
\pic {comfort goal};
\pic [left=3 of comfort goal] {energy goals};
\draw[goalrelation] (comfort goal) -- (lowerenergy goal) node[midway,above,] {\scriptsize Lower e $\Leftrightarrow$ Lower t};
\pic [right=1 of comfort goal] {freezeprotect goal={highlight}};

\begin{scope}[xshift=-3cm,yshift=-6cm] 
    \begin{scope}[on behind layer]
       \node[system,fill=gray!5,label={[below]north:{\textbf{twin system} Freeze Protected Thermostat}},minimum width=13.5cm,minimum height=3cm] (energysave-system) at (-.125,.25) {};
    \end{scope}
    
    \pic (energysave-dt) at (-4.5, 0) {energysave dt};
    \node[port,label={[portlabel]below:{Presence}}] (presence) at (energysave-system.south -| energysave-dt-presence) {$\uparrow$};
    \draw[flow] (presence) -- (energysave-dt-presence);

    \pic (comfort-dt) {comfort dt};
    \pic (comfort-wrap-outer) {comfort wrap ports hooked={energysave-system}};

    \draw[flow] (energysave-dt-comfort-temp-out) -- node[portlabel,above,midway,yshift=.5em]{User Comfort\\Temperature} (comfort-dt-comfort-temp); 
    
    \pic (freezeprotect-dt) at (4.5, 0) {freezeprotect dt={highlight}};
    \draw[flow,highlight] (freezeprotect-dt-heater.south) -- (comfort-wrap-outer-heater.north); 
    \draw[flow,highlight] (comfort-wrap-outer-room-temp.north) -- (freezeprotect-dt-temp.west);

\end{scope}

\draw[flow] (comfort-wrap-outer-comfort-temp) -| ($(comfort-wrap-outer-comfort-temp)!.5!(energysave-dt-comfort-temp-in)$) |- (energysave-dt-comfort-temp-in);

\draw[relation] (comfort-dt-dt) -- (comfort goal);
\draw[relation] (energysave-dt-dt) -- (whenabsent goal);
\draw[relation,highlight] (freezeprotect-dt-dt) -- (freezeprotect goal);

\pic {frameSeparation={whenabsent goal}};  
\end{tikzpicture}
}
\caption{Orthogonal Freeze Protection DarTwin introduces the goal \textsf{No Freezing} which is satisfied by the \textbf{Freeze Protection} \DT.}
\label{fig:freeze_protect}
\end{figure}

Initially, we deployed the \textsf{Freeze Protection} \DT as an orthogonal \DT within the twin system, but still deployed on the same Raspberry Pi and applying the same sensors and actuator. This is depicted in \Cref{fig:freeze_protect}.

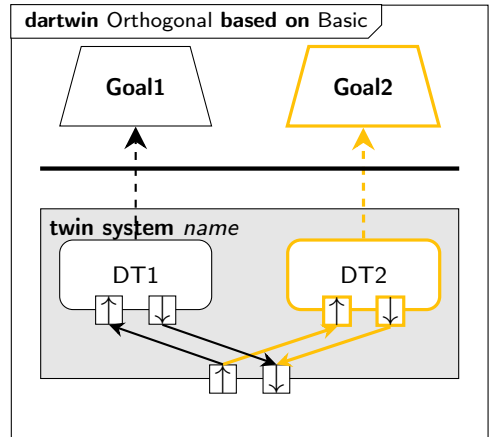
\begin{figure}
\centering
\begin{tikzpicture}[dartwinFramed={Orthogonal}{Basic}]
\pic (goal1) at (-3,0) {schema goal={Goal1/}};
\pic [right=1.25 of goal1-goal] (goal2) {schema goal={Goal2/highlight}};

\begin{scope}[yshift=-2.5cm]  
    \begin{scope}[on behind layer]
       \node[system,fill=gray!20,label={[below right]north west:{\scriptsize \textbf{twin system} \textit{name}}},minimum width=5.5cm,minimum height=2.25cm] (system2) at (-1.5cm,-.25cm) {};
    \end{scope}

    \pic (schema-dt1) at (-3, 0){schema dt={DT1/}};
    \pic (schema-dt2) at (0, 0){schema dt={DT2/highlight}};
    
    \pic (system-inout) {schema inout={system2/}};

    \draw[flow,highlight] (system-inout-in.north) -- (schema-dt2-in.south);
    \draw[flow,highlight] (schema-dt2-out.south) -- (system-inout-out.north);

    \draw[flow] (system-inout-in.north) -- (schema-dt1-in.south);
    \draw[flow] (schema-dt1-out.south) -- (system-inout-out.north);

\end{scope}

\draw[relation] (schema-dt1-dt) -- (goal1-goal);
\draw[relation,highlight] (schema-dt2-dt) -- (goal2-goal);

\pic {frameSeparation={goal}};  
\end{tikzpicture}
\caption{Orthogonal transformation}
\label{fig:schema-generalized}
\end{figure}

Watchdogs are commonly used in many \acp{cps} such as elevators, assembly lines and trains. In critical systems, there may even be several watchdogs serving the same purpose. The orthogonal transformation summarizes such a setup in \Cref{fig:schema-generalized}. The essence of a watchdog is that it is orthogonal and independent of the main functionality, taking care of the exceptional cases with unfortunate consequences.

However, with this orthogonal architecture, we realize that both the \textsf{Thermostat Logic} \DT and the \textsf{Freeze Protection} \DT may try and instruct the heat actuator at the same time, and thus, in principle, they may be in conflict. One could imagine a hacker being able to set the comfort temperature to minus 5 degrees Celsius which would obviously counter the purpose of the \textsf{Freeze Protection} \DT. Safety precautions are important, but it is necessary to consider the situations where there are competing actuations. Obviously, one could prevent such conflicts by limiting the legal domain of comfort temperatures to what is above the \textsf{No Freezing} goal boundaries, but that would make the existing \DT behaviour dependent upon the introduced watchdog. We present two alternative architectures for avoiding the potential actuation conflict.

\subsubsection{Evolution 2b: Freeze Protection with new output transformation}

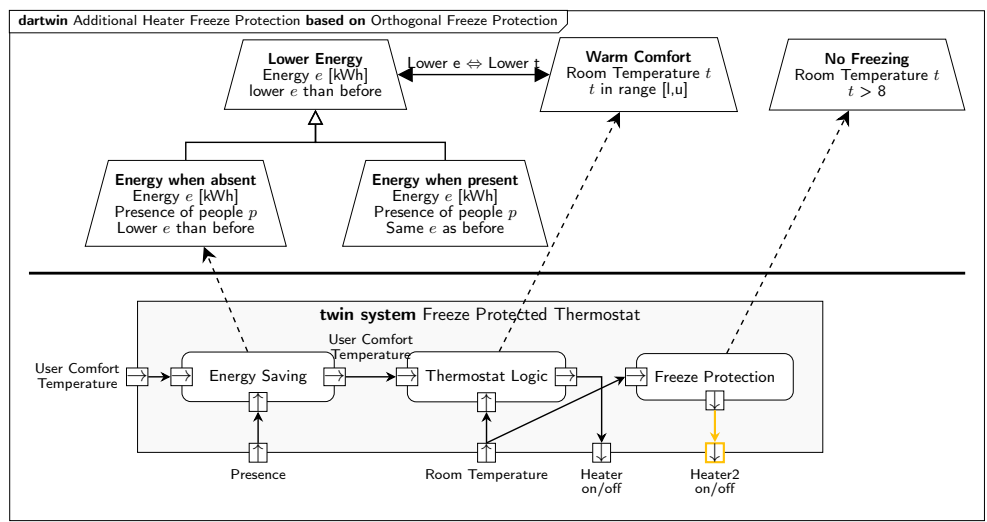
\begin{figure}[htb]
\centering
\resizebox{\textwidth}{!}{
\begin{tikzpicture}[dartwinFramed={Additional Heater Freeze Protection}{Orthogonal Freeze Protection}]
\pic {comfort goal};
\pic [left=3 of comfort goal] {energy goals};
\draw[goalrelation] (comfort goal) -- (lowerenergy goal) node[midway,above,] {\scriptsize Lower e $\Leftrightarrow$ Lower t};
\pic [right=1 of comfort goal] {freezeprotect goal};

\begin{scope}[xshift=-3cm,yshift=-6cm] 
    \begin{scope}[on behind layer]
       \node[system,fill=gray!5,label={[below]north:{\textbf{twin system} Freeze Protected Thermostat}},minimum width=13.5cm,minimum height=3cm] (energysave-system) at (-.125,0) {};
    \end{scope}
    
    \pic (energysave-dt) at (-4.5, 0) {energysave dt};
    \node[port,label={[portlabel]below:{Presence}}] (presence) at (energysave-system.south -| energysave-dt-presence) {$\uparrow$};
    \draw[flow] (presence) -- (energysave-dt-presence);

    \pic (comfort-dt) {comfort dt};
    \pic (comfort-wrap-outer) {comfort wrap ports hooked={energysave-system}};

    \draw[flow] (energysave-dt-comfort-temp-out) -- node[portlabel,above,midway,yshift=.5em]{User Comfort\\Temperature} (comfort-dt-comfort-temp); 
    
    \pic (freezeprotect-dt) at (4.5, 0) {freezeprotect dt};

    \node[port,highlight,label={[portlabel]below:{Heater2 \\on/off}}] (heater2) at (freezeprotect-dt-heater |- energysave-system.south) {$\downarrow$};
    \draw[flow,highlight] (freezeprotect-dt-heater.south) -- (heater2.north); 
    
    \draw[flow] (comfort-wrap-outer-room-temp.north) -- (freezeprotect-dt-temp.west);

\end{scope}

\draw[flow] (comfort-wrap-outer-comfort-temp) -| ($(comfort-wrap-outer-comfort-temp)!.5!(energysave-dt-comfort-temp-in)$) |- (energysave-dt-comfort-temp-in);

\draw[relation] (comfort-dt-dt) -- (comfort goal);
\draw[relation] (energysave-dt-dt) -- (whenabsent goal);
\draw[relation] (freezeprotect-dt-dt) -- (freezeprotect goal);

\pic {frameSeparation={whenabsent goal}};  

\end{tikzpicture}
}
\caption{Additional Heater Freeze Protection DarTwin removes the actuator conflict by introducing an additional heater output.}
\label{fig:freeze_protect_noconflict}
\end{figure}

The problem with the purely orthogonal architecture is that the single heater may be independently instructed and these instructions may be conflicting. One solution would be to have two heaters, one instructed by the \textsf{Thermostat Logic} \DT and one controlled by the \textsf{Freeze Protectoin} \DT. The freeze protect heater (\textsf{Heater 2} in \Cref{fig:freeze_protect_noconflict}) must be strong enough to make the room not freeze, but it does not necessarily need to be strong enough to maintain the comfort temperature. This is a good solution (\Cref{fig:freeze_protect_noconflict}) that copes with situations where the original heater is not functioning, or the thermostat logic is erroneous.

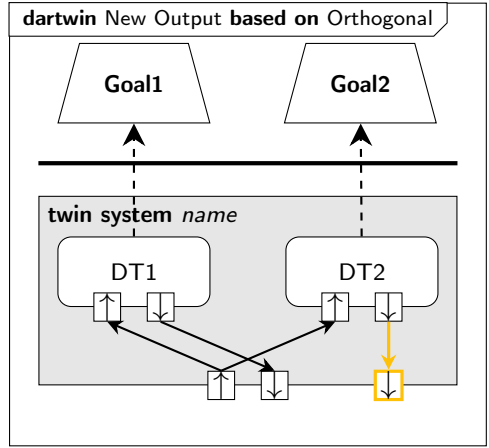
\begin{figure}
\centering
\begin{tikzpicture}[dartwinFramed={New Output}{Orthogonal}]
\pic (goal1) at (-3,0) {schema goal={Goal1/}};
\pic [right=1.25 of goal1-goal] (goal2) {schema goal={Goal2/}};


\begin{scope}[yshift=-2.5cm]  
    \begin{scope}[on behind layer]
       \node[system,fill=gray!20,label={[below right]north west:{\scriptsize \textbf{twin system} \emph{name}}},minimum width=5.5cm,minimum height=2.5cm] (system2) at (-1.5cm,-.25cm) {};
    \end{scope}

    \pic (schema-dt1) at (-3, 0){schema dt={DT1/}};
    \pic (schema-dt2) at (0, 0){schema dt={DT2/}};
    
    \pic (system-inout) {schema inout={system2/}};
    
    \node[port,highlight,label={below:{\scriptsize }}] (system-out) at (system2.south -| schema-dt2-out) {$\downarrow$};

    \draw[flow] (system-inout-in.north) -- (schema-dt1-in.south);
    \draw[flow] (system-inout-in.north) -- (schema-dt2-in.south);
    \draw[flow] (schema-dt1-out.south) -- (system-inout-out.north);
    \draw[flow,highlight] (schema-dt2-out.south) -- (system-out.north);
    
\end{scope}

\draw[relation] (schema-dt1-dt) -- (goal1-goal);
\draw[relation] (schema-dt2-dt) -- (goal2-goal);

\pic {frameSeparation={goal1-goal}};  
\end{tikzpicture}
\caption{New Output transformation.}
\label{fig:schema-new-output}
\end{figure}

In general the transformation template given in \Cref{fig:schema-new-output} shows how the watchdog could be defined by making the control data separate to two distinct actuators, while receiving monitoring data in both \DTs is normally not provoking a conflict.

\subsubsection{Evolution 2c: Freeze Protection with chained transformation} 

Our real system has only one heater, and therefore we look for an alternative solution.
We would rather let the architecture reflect the desired priority, namely that the watchdog should have priority since it will be guarding the situation with the most serious consequences. We can achieve that by modifying the \DT architecture to obtain a pipelined structure.
We would like to transform the architecture such that one DT (\textsf{Freeze Protection}) is given priority over another \DT (\textsf{Thermostat Logic}) relative to the actuator (\textsf{Heater}). This is depicted in \Cref{fig:freeze_protect_prio} where we see that actuation from the \textsf{Thermostat Logic} will pass through \textsf{Freeze Protection} and one may say that \textsf{Freeze Protection} assesses and filters the control data from the \textsf{Thermostat Logic} against the freeze conditions (temperature $t > 8^{\circ}C$).

\begin{figure}
\centering
\begin{tikzpicture}[dartwinFramed={Chaining}{Orthogonal}]
\pic (goal1) at (-3,0) {schema goal={Goal1/}};
\pic [right=1.25 of goal1-goal] (goal2) {schema goal={Goal2/}};

\begin{scope}[yshift=-2.5cm]  
    \begin{scope}[on behind layer]
       \node[system,fill=gray!20,label={[below right]north west:{\scriptsize \textbf{twin system} \textit{name}}},minimum width=5.5cm,minimum height=2.25cm] (system2) at (-1.5cm,-.25cm) {};
    \end{scope}
    \pic (system-inout) {schema inout={system2/}};

    \pic (schema-dt1) at (-3, 0){schema dt only={DT1/}};
    \node[port,label={below:{\scriptsize }}] (dt1-right-out) at (schema-dt1-dt.east) {$\rightarrow$};
    \node[port,label={below:{\scriptsize }},xshift=-1em] (dt1-south-in) at (schema-dt1-dt.south) {$\uparrow$};
    
    \pic (schema-dt2) at (0, 0){schema dt={DT2/}};
    \node[port,highlight,label={below:{\scriptsize }}] (dt2-left-in) at (schema-dt2-dt.west) {$\rightarrow$};

\draw[flow] (system-inout-in.north) -- (schema-dt1-in.south);
\draw[flow] (system-inout-in.north) -- (schema-dt2-in.south);
\draw[flow] (schema-dt2-out.south) -- (system-inout-out.north);

\draw[flow,highlight] (dt1-right-out) -- (dt2-left-in);

\end{scope}

\draw[relation] (schema-dt1-dt) -- (goal1-goal);
\draw[relation] (schema-dt2-dt) -- (goal2-goal);

\pic {frameSeparation={goal1-goal}};  
\end{tikzpicture}
\caption{Chaining transformation.}
\label{fig:schema-chaining}
\end{figure}
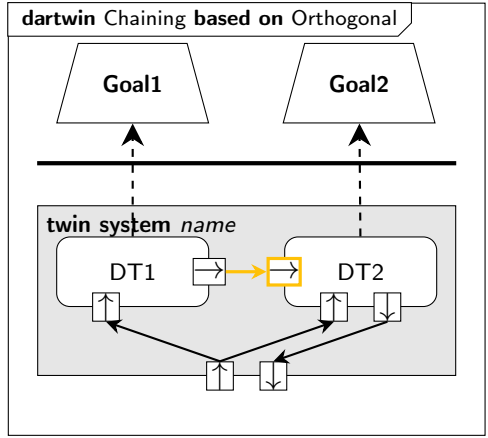

Our aim was to achieve priority of the watchdog, and this was provided by piping the actuation. We refer to this transformation pattern as ``chaining'' and in this case priority is achieved by ``post-chaining'' the \DT. We may say that the \textsf{Energy Saving} \DT is ``pre-chaining'' the \textsf{Thermostat Logic} for its purpose of creating the two sub-cases of \textsf{Warm Comfort}. The Chaining transformation is indicated in \Cref{fig:schema-chaining}.

The chained \DT architecture avoids actuation conflict by applying priorities, but fails to prevent consequences of a failing heater. This watchdog merely covers thermostat logic errors.

Our real implemented system also has one more watchdog that monitors the opposite situation: when the heater is on for so long that it may cause fire. We have pipelined the \textsf{Fire Protection} watchdog following \textsf{Freeze Protection} giving fire a higher priority than freezing since a fire has even more consequences than frozen pipes. Note that, since there are no new considerations regarding the pipelined \textsf{Fire Protection} watchdog, we omitted its description in this paper. The principle behind \textsf{Fire Protection} is to make sure that the heater cools off after a given maximum operating period. This duration is of course a parameter.

\begin{figure}[htb]
\centering
\resizebox{\textwidth}{!}{
\begin{tikzpicture}[dartwinFramed={Chained Freeze Protection}{Orthogonal Freeze Protection}]
\pic {comfort goal};
\pic [left=3 of comfort goal] {energy goals};
\draw[goalrelation] (comfort goal) -- (lowerenergy goal) node[midway,above,] {\scriptsize Lower e $\Leftrightarrow$ Lower t};
\pic [right=1 of comfort goal] {freezeprotect goal};

\begin{scope}[xshift=-3cm,yshift=-6cm] 
    \begin{scope}[on behind layer]
       \node[system,fill=gray!5,label={[below]north:{\textbf{twin system} Freeze Protected Thermostat}},minimum width=13.5cm,minimum height=3.25cm] (energysave-system) at (-.125,0) {};
    \end{scope}

    \pic (energysave-dt) at (-4.5, 0) {energysave dt};
    \node[port,label={below:{\tiny Presence}}] (presence) at (energysave-system.south -| energysave-dt-presence) {$\uparrow$};
    \draw[flow] (presence) -- (energysave-dt-presence);

    \pic (comfort-dt) {comfort dt};
    \pic (comfort-wrap-outer) {comfort wrap ports={energysave-system}};

    \draw[flow] (comfort-wrap-outer-room-temp) -- (comfort-dt-temp);
    
    \draw[flow] (energysave-dt-comfort-temp-out) --node[portlabel,above,midway,yshift=.5em]{User Comfort\\Temperature} (comfort-dt-comfort-temp); 

    \node[process,minimum height=1.5cm] (freezeprotect-dt-dt) at (4.75,-.5em) {Freeze Protection};
    \node[port,label={below:{\scriptsize }},yshift=-1em] (freezeprotect-dt-temp) at (freezeprotect-dt-dt.west) {$\rightarrow$};
    \node[port,highlight,label={below:{\scriptsize }},yshift=.5em] (freezeprotect-dt-heater-in) at (freezeprotect-dt-dt.west) {$\rightarrow$};
    \node[port,label={below:{\scriptsize }}] (freezeprotect-dt-heater) at (freezeprotect-dt-dt.south) {$\downarrow$};
    
    \draw[flow] (freezeprotect-dt-heater.south) -- (comfort-wrap-outer-heater.north); 
    \draw[flow] (comfort-wrap-outer-room-temp.north) -- (freezeprotect-dt-temp.west);

    \draw[flow,highlight] (comfort-dt-heater) -- node[portlabel,above,midway,yshift=.5em]{Temperature} (freezeprotect-dt-heater-in);
    
\end{scope}

\draw[flow] (comfort-wrap-outer-comfort-temp) -| ($(comfort-wrap-outer-comfort-temp)!.5!(energysave-dt-comfort-temp-in)$) |- (energysave-dt-comfort-temp-in);

\draw[relation] (comfort-dt-dt) -- (comfort goal);
\draw[relation] (energysave-dt-dt) -- (whenabsent goal);
\draw[relation] (freezeprotect-dt-dt) -- (freezeprotect goal);

\pic {frameSeparation={whenabsent goal}};  

\end{tikzpicture}
}
\caption{Chained Freeze Protection DarTwin applies the chaining transformation to resolve the actuator conflict.}
\label{fig:freeze_protect_prio}
\end{figure}
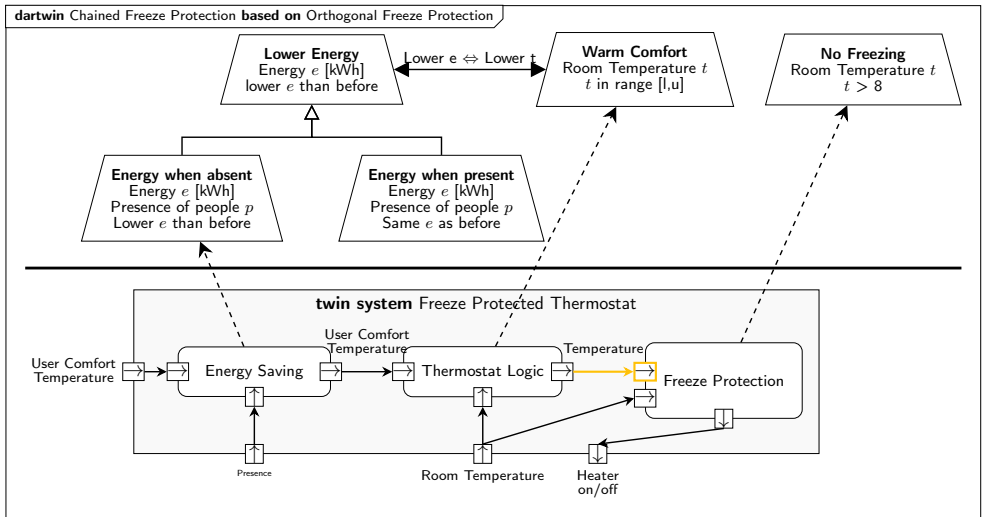

\subsubsection{Evolution 3: Saving Money with arbitration transformation}
\label{sec:arbiter-intro}

In 2022 the electricity prices in Norway changed drastically. The price of electrical energy soared to 200-500\% and became unpredictably volatile. The reasons were the combinations of a steep increase in foreign demand due to the Ukrainian war, and some new large cables to bring the energy to Europe. Suddenly, the price of energy in Norway was dependent on the energy prices in Germany. 
Thus, next to energy saving, which also impacts the amount of money, the hourly price variations became an even more important factor than the amount of energy used.

With this background, we first redesigned our \DT on energy. We replaced the \textsf{Energy Saving} \DT that was based on lower temperature when absent, with a process that focused on energy cost. 
The applied approach was simplistic: We assume that the inhabitants would accept lower temperatures when the energy prices were high. Therefore, we devised a function (implemented by a lookup table) that reads the hourly varying price as input and provided a delta temperature value of how much the inhabitant was willing to lower the comfort temperature (and instead rather put on a jumper or a woollen blanket). Thus, the replaced DT on energy-cost saving had fundamentally the same interface towards the thermostat, namely to change the comfort temperature setting and there would be no further architectural changes.

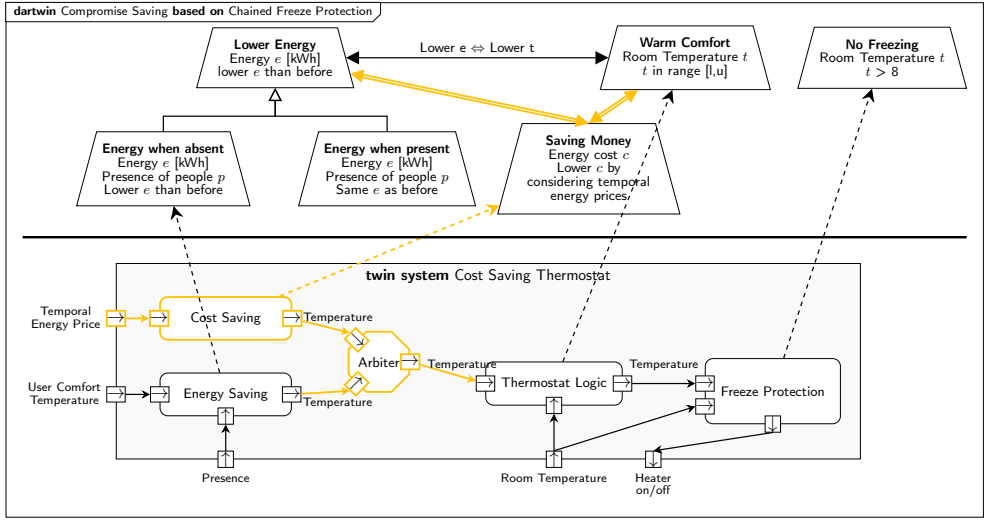
\begin{figure}[htb]
\centering
\resizebox{\textwidth}{!}{
\begin{tikzpicture}[dartwinFramed={Compromise Saving}{Chained Freeze Protection}]

\pic {comfort goal};
\pic [left=6 of comfort goal] {energy goals};
\draw[goalrelation] (comfort goal) -- (lowerenergy goal) node[midway,above,] {\scriptsize Lower e $\Leftrightarrow$ Lower t};

\pic [right=1 of whenpresent goal]{savemoney goal};
\draw[conflict,highlight] (savemoney goal.north) -- (lowerenergy goal);
\draw[conflict,highlight] (savemoney goal.north) -- (comfort goal);

\pic [right=1 of comfort goal] {freezeprotect goal};


\begin{scope}[xshift=-3cm,yshift=-7.5cm] 
    \begin{scope}[on behind layer]
       \node[sys,fill=gray!5,label={[below]north:{\textbf{twin system} Cost Saving Thermostat}},minimum width=17cm,minimum height=4.5cm] (energysave-system) at (-1.5,.25) {};
    \end{scope}

    \pic (energysave-dt) at (-7.5, -.25) {energysave dt};
    \node[port,label={[portlabel]below:{Presence}}] (presence) at (energysave-system.south -| energysave-dt-presence) {$\uparrow$};
    \draw[flow] (presence) -- (energysave-dt-presence);

    \pic (costsave-dt) at (-7.5, 1.5) {costsave dt={highlight}};
    \pic (arbiter-dt) at (-4, 0.5) {arbiter dt={highlight}};

    \pic (comfort-dt) {comfort dt};
    \pic (temp-heat-outer) {temp heat ports={energysave-system}};

    \draw[flow] (temp-heat-outer-room-temp) -- (comfort-dt-temp);
    \draw[flow,highlight] (arbiter-dt-comfort-temp) -- node[portlabel,above,pos=0.75]{Temperature} (comfort-dt-comfort-temp);
    
    \draw[flow,highlight] (costsave-dt-comfort-temp) -- node[portlabel,above,pos=0.75]{Temperature} (arbiter-dt-in1.west); 
    \draw[flow,highlight] (energysave-dt-comfort-temp-out) -- node[portlabel,below,pos=0.75]{Temperature} (arbiter-dt-in2.west); 

    \node[port,highlight,label={[portlabel]left:{Temporal\\Energy Price}}] (energysave-price-in) at (energysave-system.west |- costsave-dt-price) {$\rightarrow$};

    \draw[flow,highlight] (energysave-price-in) -- (costsave-dt-price); 

    \node[process,minimum height=1.5cm] (freezeprotect-dt-dt) at (5,-.5em) {Freeze Protection};
    \node[port,label={below:{}},yshift=-1em] (freezeprotect-dt-temp) at (freezeprotect-dt-dt.west) {$\rightarrow$};
    \node[port,label={below:{}},yshift=.5em] (freezeprotect-dt-heater-in) at (freezeprotect-dt-dt.west) {$\rightarrow$};
    \node[port,label={below:{}}] (freezeprotect-dt-heater) at (freezeprotect-dt-dt.south) {$\downarrow$};
    
    \draw[flow] (freezeprotect-dt-heater.south) -- (temp-heat-outer-heater.north); 
    \draw[flow] (temp-heat-outer-room-temp.north) -- (freezeprotect-dt-temp.west);

    \draw[flow] (comfort-dt-heater) -- node[portlabel,above,midway,yshift=.5em]{Temperature} (freezeprotect-dt-heater-in);
    
\end{scope}

\node[port,label={[portlabel]left:{User Comfort\\Temperature}}] (-comfort-temp) at (energysave-system.west |- energysave-dt-comfort-temp-in) {$\rightarrow$};
\draw[flow] (-comfort-temp) -- (energysave-dt-comfort-temp-in);

\draw[relation] (comfort-dt-dt) -- (comfort goal);
\draw[relation] (energysave-dt-dt) -- (whenabsent goal);
\draw[relation] (freezeprotect-dt-dt) -- (freezeprotect goal);
\draw[relation,highlight] (costsave-dt-dt) -- (savemoney goal);

\pic {frameSeparation={savemoney goal}};  

\end{tikzpicture}
}
\caption{Compromise Saving DarTwin introduces the \textsf{Saving Money} goal, which is fulfilled by the \textsf{Cost Saving} \DT in combination with the \textsf{Arbiter}.}
\label{fig:arbiter}
\end{figure}

However, the original idea of changing the comfort temperature when there is nobody present, still seems a good idea even when we have introduced the new cost-centric approach. Could we apply both strategies together? If so, how should they be combined? We have now two \DTs concerned with energy that both try to manipulate the comfort temperature. Therefore there is a potential conflict between the goals to save money and to lower energy. This is indicated by a double arrow as seen in \Cref{fig:arbiter}. We have defined a conflict arrow between \textsf{Warm Comfort} and \textsf{Saving Money} as well since our approach is to sacrifice some comfort for money.

What we have now are two \DTs that each suggests a new comfort temperature. What is needed is some explicit arbitration facility for the combination of the two suggestions from the \DTs involved with energy. One arbitration rule could be to choose the lowest of the two suggestions at all times. This arbitration architecture is depicted in \Cref{fig:arbiter}. It shows that our additional goal now is to save money, and our operationalization is to take advantage of knowing the hourly changing energy prices. In our implementation, the hourly cost is given as a table which gets its values from a public website depicted in our illustration as a port connected to an external API giving input to the \textsf{Cost Saving} \DT.

\begin{figure}
\centering
\begin{tikzpicture}[dartwinFramed={Arbitration}{New Output}]
\pic (goal1) at (-3,0) {schema goal={Goal1/}};
\pic [right=1.25 of goal1-goal] (goal2) {schema goal={Goal2/}};

\begin{scope}[yshift=-2.5cm]  
    \begin{scope}[on behind layer]
       \node[system,fill=gray!20,label={[below right]north west:{\scriptsize \textbf{twin system} \textit{name}}},minimum width=5.5cm,minimum height=4.25cm] (system2) at (-1.5cm,-1cm) {};
    \end{scope}

    \pic (schema-dt1) at (-3, 0){schema dt={DT1/}};
    \pic (schema-dt2) at (0, 0){schema dt={DT2/}};
    \pic (schema-arbiter) at (-.25, -1.75){schema arbiter dt={Arbiter/highlight}};
    \draw[flow,highlight] (schema-dt1-out.south) -- (schema-arbiter-in1.north);
    \draw[flow,highlight] (schema-dt2-out.south) -- (schema-arbiter-in2.north);
    
    \node[port,label={below:{\scriptsize }}] (system-south-in) at (schema-dt1-in |- system2.south) {$\uparrow$};
    \draw[flow] (system-south-in) -- (schema-dt1-in);
    \draw[flow] (system-south-in) -- (schema-dt2-in);

    \node[port,label={below:{\scriptsize }}] (system-south-out) at (schema-arbiter-out |- system2.south) {$\downarrow$};
    \draw[flow,highlight] (schema-arbiter-out) -- (system-south-out);
    
\end{scope}

\draw[relation] (schema-dt1-dt) -- (goal1-goal);
\draw[relation] (schema-dt2-dt) -- (goal2-goal);

\pic {frameSeparation={goal1-goal}};  

\end{tikzpicture}
\caption{Arbitration transformation.}
\label{fig:schema-arbiter}
\end{figure}
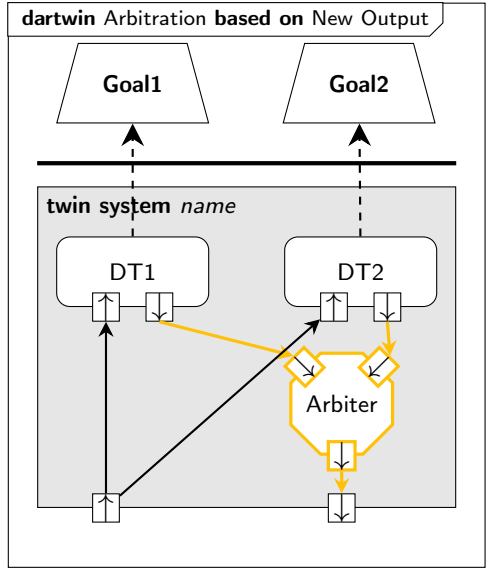

The arbiter architecture as depicted in \Cref{fig:schema-arbiter} is well-known from \eg safety-critical systems such as train monitoring systems where there are two totally independent watchdogs with the same requirement, but with fundamentally different implementations with disjunct development teams, disjunct hardware and disjunct software. The arbiter's role is to initiate an emergency exception if the two watchdogs do not agree. In our situation, the arbiter's role is also to harmonize and combine the results from the two energy DTs. Our first implementation of this harmonization is very simple, but it represents the process of compromise, and it may be enhanced to any kind of complexity. 

\section{Evaluation}\label{sec:evaluation}

\begin{figure}
    \centering
    \includegraphics[width=0.7\textwidth]{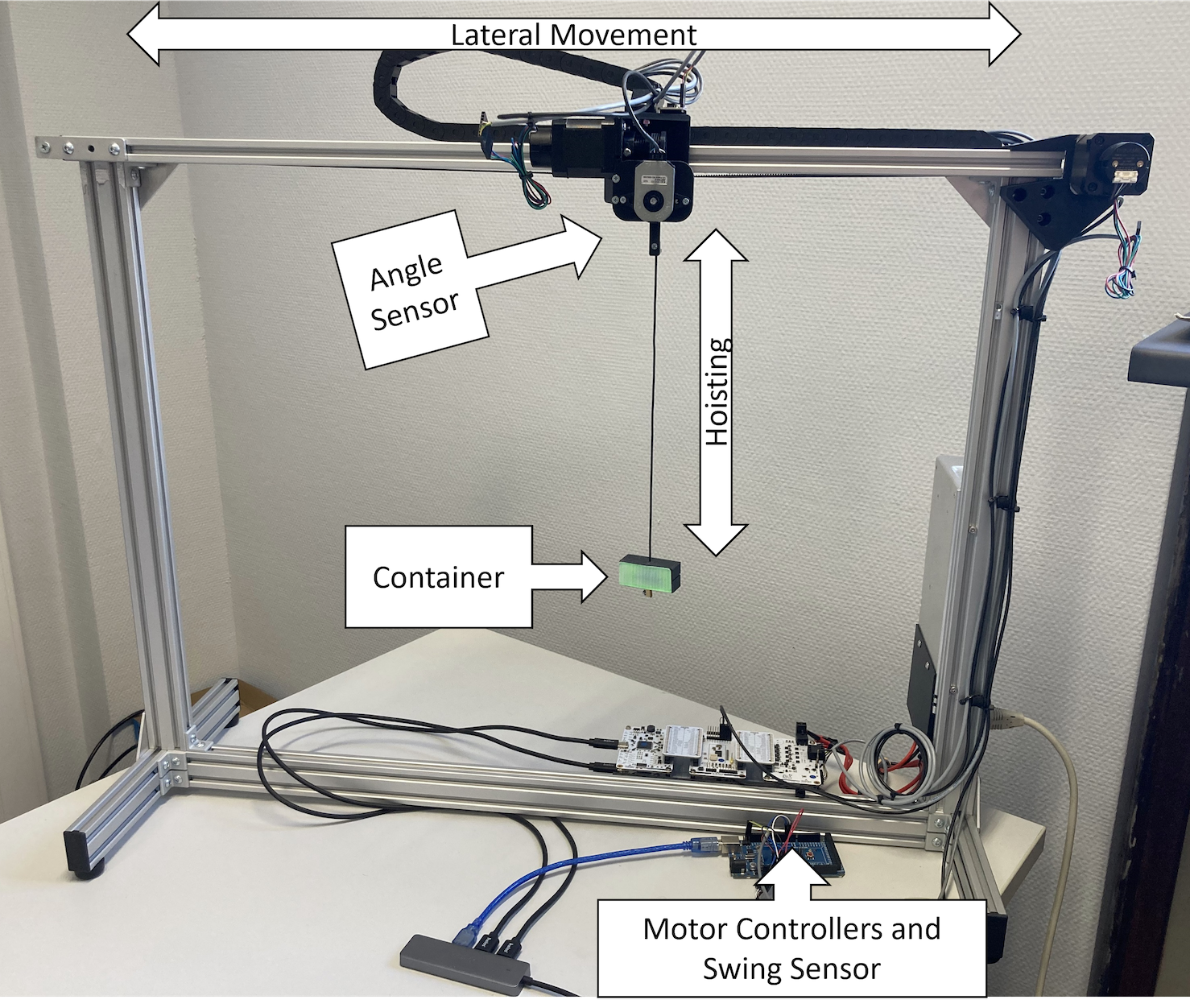}
    \caption{Picture of the lab-scale gantry crane.}
    \label{fig:real-gc}
\end{figure}

So far, we explained the transformations with the help of the smart home system. Now, we want to evaluate whether or not the transformations are also applicable to systems in other domains. To this end, we apply them (retroactively) to another twin system at the Univeristy of Antwerp's Cosys-Lab: a lab-scale \textsf{Gantry Crane} twin system, shown in \cref{fig:real-gc}. The goal of a real gantry crane is to move containers from/onto ships that are docked in the harbour. To achieve a high throughput, the swinging motion at the end of a container movement should be minimized, and preferably even zero. This way, the container can be lowered onto the quay or ship immediately and safely. Our lab-scale crane is a scaled down (approximately 1:10) version of such a real gantry crane, moving a container of 5 cm x 2 cm x 2 cm (about 1:10 of a 20' standard container) over a height and width of about 50 x 70 cm. Suffice to say, our lab-scale crane does not move the containers onto real ships, but it does move a container payload from one position to another, mimicking such a real crane. We discuss 4 stages in the development of the crane twin system, each one incrementally adds to the previous stage:

\begin{itemize}
    \item The initial \textsf{Gantry Crane} twin system (\cref{sec:optimalcontroltwinsystem}): the container is moved over an optimally calculated trajectory. This stage shows the use of the \textsf{Basic} DarTwin from \cref{fig:schema-new-goal}.
    \item Evolution 1 - Dynamic Positioning Constraints (\cref{sec:dynamicpositioningconstraints}): we add dynamic positioning constraints to the \textsf{Gantry Crane} twin system. These constraints form dynamic keep-out zones that influence the calculated trajectory. This stage shows the \textsf{Augmented} transformation with multiple \DTs from \cref{fig:schema-flat}.
    \item Evolution 2 - Continuous Validation (\cref{sec:continuousvalidation}): after each executed trajectory, a validation is performed, checking that the trajectory was executed flawlessly (or not). This stage shows a slightly modified \textsf{New Output} transformation from \cref{fig:schema-new-output}.
    \item Evolution 3 - Container Constraints(\cref{sec:containerconstraints}): dynamic kinetic constraints based on the container's contents are added to influence the calculated trajectory. This stage shows an alternative way to add dynamic constraints based on the \textsf{Arbiter} transformation from \cref{fig:schema-arbiter}.
\end{itemize}

\subsection{Optimal Control Twin System}
\label{sec:optimalcontroltwinsystem}

\begin{figure}[htb]
\centering
\resizebox{.5\textwidth}{!}{
\begin{tikzpicture}[dartwinFramed={Optimal Control}{}]

\pic {constraints goal};
\pic [below left=of constraints goal] {noswing goal};
\pic [below right=of constraints goal] {minimize goal};

\begin{scope}[yshift=-5.5cm]  
    \pic (trajectory-dt) {trajectory dt};
    \begin{scope}[on behind layer]
       \node[sys,fill=gray!20,label={[below]north:{\scriptsize \textbf{twin system} Gantry Crane}},minimum width=5cm,minimum height=2.5cm] (system) at (0,-.5) {};
    \end{scope}
    \pic (trajectory-wrap) {wrap trajectory ports={system}};
\end{scope}

\draw[relation] (trajectory-dt-dt) -- (constraints goal);
\draw[relation] (trajectory-dt-dt) -- (noswing goal);
\draw[relation] (trajectory-dt-dt) -- (minimize goal);

\pic {frameSeparation={minimize goal}};  
\end{tikzpicture}
}
\caption{Optimal Control DarTwin showing the initial \textsf{Gantry Crane} twin system.}
\label{fig:gantry-crane-initial}
\end{figure}
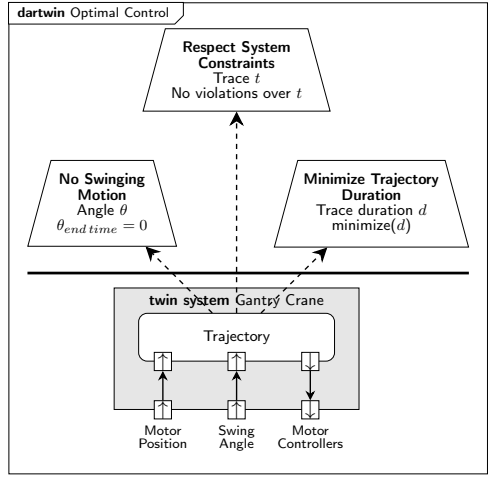

\cref{fig:gantry-crane-initial} shows the initial \textsf{Optimal Control} twin system. It showcases the \textsf{Basic} DarTwin from \cref{fig:schema-new-goal}. The goal of the crane is to achieve a high throughput of containers. This is achieved by following an optimal trajectory, generated by a trajectory generation service provided by the \textsf{Trajectory} \DT. Every time the crane should move from one point to another, one such trajectory must be calculated. The \textsf{Trajectory} \DT contains an optimal constraint problem solver, which tries to find an optimal trajectory that the crane can execute to move the container. The goals of this \DT are threefold: (i) there should be no swinging motion at the end of the trajectory, (ii) the duration of the trajectory should be as short as possible, and (iii) none of the system constraints may be violated. For the goal that minimizes the trajectory duration, the \Ac{poi} is the time duration in seconds of the trajectory. This duration should be minimized. For the goal of having no swinging motion, the \ac{poi} are swinging angle and angular velocity in rad and rad/s, both should be equal to zero at the end of the trajectory (the latter is not shown in the figure for conciseness). Lastly, for the adherence to the system constraints goal, we have both a set of geometric constraints and kinetic constraints to which the \textsf
{Trajectory} \DT should adhere. For the geometric constraints, the \Ac{poi} are the minimum and maximum positions of the cart and the container. For the kinetics, the \Ac{poi} are the maximum accelerations and velocities of the cart and container. These properties stem from the geometry of the crane, as well as the properties of the motors driving the crane, and should be strictly adhered to. To generate the trajectory, the \textsf{Trajectory} \DT uses the current motor position and swing angle as initial conditions/inputs, and yields an optimal trajectory that can be executed by the motor controllers as output. Once generated, the motor controllers accept this trajectory and enact it in real life. Implementation-wise, this means the motor controllers set the end position of the trajectory as target, then, as the cart is moving along the trajectory, they update the cart's velocity over time, as specified by the samples in the trajectory.

\subsection{Evolution 1: Dynamic Positioning Constraints}
\label{sec:dynamicpositioningconstraints}

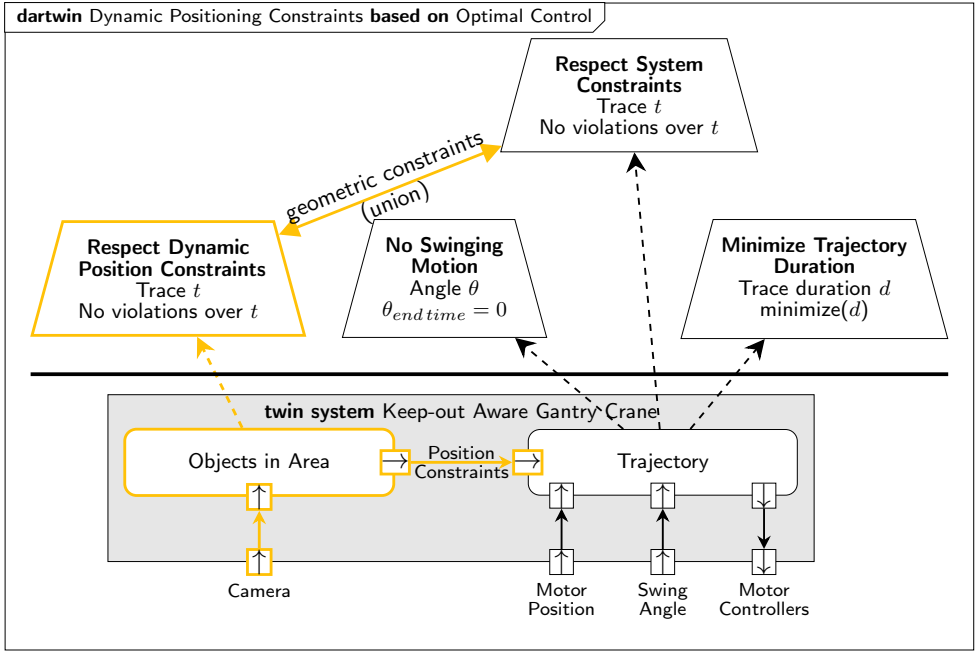
\begin{figure}[htb]
\centering
\resizebox{\textwidth}{!}{
\begin{tikzpicture}[dartwinFramed={Dynamic Positioning Constraints}{Optimal Control}]

\pic {constraints goal};
\pic [below left=of constraints goal] {noswing goal};
\pic [below right=of constraints goal] {minimize goal};
\pic [left=of noswing goal] {nocollision goal={highlight}};
\draw[goalrelation,highlight] (nocollision goal) -- (constraints goal) node[midway,sloped] {geometric constraints\\(union)};

\begin{scope}[yshift=-5.5cm]  
    \pic (trajectory-dt) at (.5,0) {trajectory dt};
    \begin{scope}[on behind layer]
       \node[sys,fill=gray!20,label={[below]north:{\scriptsize \textbf{twin system} Keep-out Aware Gantry Crane}},minimum width=10.5cm,minimum height=2.5cm] (system) at (-2.5,-.5) {};
    \end{scope}
    \pic (trajectory-wrap) {wrap trajectory ports={system}};

    \pic (objectsarea-dt) at (-5.5,0) {objectsarea dt={highlight}};
    \node[port,highlight,label={[portlabel]below:{Camera}}] (system-camera) at (system.south -| objectsarea-dt-camera) {$\uparrow$};
    \draw[flow,highlight] (system-camera) -- (objectsarea-dt-camera);

\end{scope}

\node[port,highlight,label={below:{}}] (-constraints) at (trajectory-dt-dt.west) {$\rightarrow$};
\draw[flow,highlight] (objectsarea-dt-constraints) -- node[portlabel,midway] {Position\\Constraints} (-constraints);

\draw[relation] (trajectory-dt-dt) -- (constraints goal);
\draw[relation] (trajectory-dt-dt) -- (noswing goal);
\draw[relation] (trajectory-dt-dt) -- (minimize goal);
\draw[relation,highlight] (objectsarea-dt-dt) -- (nocollision goal);

\pic {frameSeparation={minimize goal}};  

\end{tikzpicture}
}
\caption{DarTwin \textsf{Dynamic Positioning Constraints} adds the goal \textsf{Respect Dynamic Position Constraints}, which is fulfilled by the \textsf{Objects in Area} \DT.}
    \label{fig:gantry-crane-evolution1}
\end{figure}

A harbour terminal is a dynamic environment, with straddle carriers moving containers in and out, deckhands securing containers on the ship and so on. In an attempt to increase the safety of the \textsf{Gantry Crane} twin system, we are looking to add dynamic keep-out zones, zones in which the container should not pass in its trajectory for safety reasons. The new goal is thus to respect the dynamic position constraints of those keep-out zones. Because these keep-out zones change over time, before a trajectory is calculated, the latest zones must be taken into account. The \Ac{poi} is the container's position trace, which at no point should violate the keep-out zones. Implementation-wise, these position constraints are additional geometric constraints that the \textsf{Trajectory} \DT must adhere to. We show this by drawing a relation between the \textsf{Respect Dynamic Position Constraints} and \textsf{Respect System Constraints} goals. Implementation-wise, the \textsf{Trajectory} \DT must adhere to the union of these goals. To achieve this goal, we must undertake two steps. Firstly, the \textsf{Trajectory} \DT must be mutated, such that it gains an input port on which it can accept these dynamic position constrains, and it must apply the union of all geometric constraints when solving for the optimal trajectory. Secondly, we need an \textsf{Objects in Area} \DT, a \DT which captures the position of objects in real time, and extracts the position constraints from that information. It should do so based on information coming from cameras spread around the gantry crane. This is visually shown in \cref{fig:gantry-crane-evolution1}. Machine learning techniques with image segmentation can be applied in the implementation to extract the keep out zones. To add this new \DT to the system, we apply the \textsf{Augmented} twin transformation with multiple \DTs as previously shown in \cref{fig:schema-flat}. The new \DT is coupled in front of the existing \DT, as was done previously in \cref{sec:greentwinflat}. 

\subsection{Evolution 2: Continuous Validation}
\label{sec:continuousvalidation}

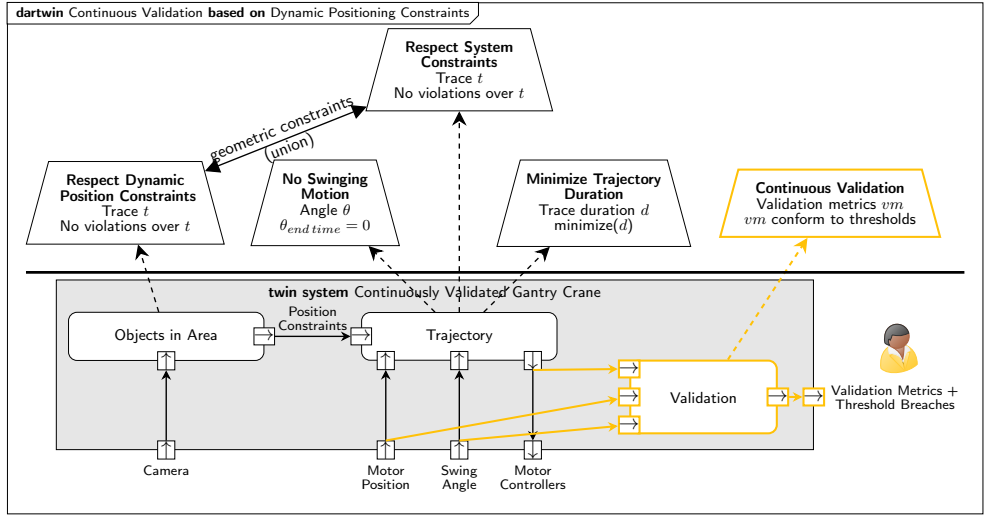
\begin{figure}[htb]
\centering
\resizebox{\textwidth}{!}{
\begin{tikzpicture}[dartwinFramed={Continuous Validation}{Dynamic Positioning Constraints}]

\pic {constraints goal};
\pic [below left=of constraints goal] {noswing goal};
\pic [below right=of constraints goal] {minimize goal};
\pic [left=of noswing goal] {nocollision goal};
\draw[goalrelation] (nocollision goal) -- (constraints goal) node[midway,sloped] {geometric constraints\\(union)};

\pic [right=of minimize goal] {validation goal={highlight}};

\begin{scope}[yshift=-5.5cm]  
    \pic (trajectory-dt) {trajectory dt};
    \begin{scope}[on behind layer]
       \node[sys,fill=gray!20,label={[below]north:{\scriptsize \textbf{twin system} Continuously Validated Gantry Crane}},minimum width=15.5cm,minimum height=3.5cm] (system) at (-.5,-.85) {};
    \end{scope}
    \pic (trajectory-wrap) {wrap trajectory ports={system}};

    \pic (objectsarea-dt) at (-6,0) {objectsarea dt};
    \node[port,label={[portlabel]below:{Camera}}] (system-camera) at (system.south -| objectsarea-dt-camera) {$\uparrow$};
    \draw[flow] (system-camera) -- (objectsarea-dt-camera);

    \pic (validation-dt) at (5,-1.25) {validation dt={highlight}};
    \node[port,highlight,label={[portlabel]right:{Validation Metrics +\\Threshold Breaches}}] (-metrics) at (validation-dt-dt.east -| system.east) {$\rightarrow$};

    \draw[flow,highlight] (trajectory-dt-controllers.south) -- (validation-dt-controllers);
    \draw[flow,highlight] (trajectory-wrap-position.north) -- (validation-dt-position);
    \draw[flow,highlight] (trajectory-wrap-swing.north) -- (validation-dt-swing);
    \draw[flow,highlight] (validation-dt-metrics) -- (-metrics);

    \node[minimum width=.75cm,minimum height=1cm,alice,shirt=NiceHighlightColor] at (9,-.25) {};
    
\end{scope}

\node[port,label={below:{}}] (-constraints) at (trajectory-dt-dt.west) {$\rightarrow$};
\draw[flow] (objectsarea-dt-constraints) -- node[midway,above,portlabel]{Position \\Constraints} (-constraints);

\draw[relation] (trajectory-dt-dt) -- (constraints goal);
\draw[relation] (trajectory-dt-dt) -- (noswing goal);
\draw[relation] (trajectory-dt-dt) -- (minimize goal);
\draw[relation] (objectsarea-dt-dt) -- (nocollision goal);
\draw[relation,highlight] (validation-dt-dt) -- (validation goal);

\pic {frameSeparation={minimize goal}};  
\end{tikzpicture}
}
\caption{DarTwin Continuous Validation adds the \textsf{Continuous Validation} goal, which is fulfilled by the \textsf{Validation} \DT.}
\label{fig:gantry-crane-evolution2}
\end{figure}

The second evolution is the addition of a continuous validation scheme. Here, we wish to continuously validate that the digital model of the gantry crane is still a good model of the physical gantry crane. This is important since the model is used in the \textsf{Trajectory} \DT to generate optimal trajectories, and set the system constraints. When this model starts deviating from the crane's actual behaviour, the generated trajectories are no longer optimal. The goal, thus, is to validate the crane model after each completed trajectory enactment. We do so by comparing every executed trajectory of the gantry crane with simulations that execute the same trajectory. This comparison produces validation metrics that should remain below a warning/failure thresholds. The properties of interest are thus the validation metrics, generated by the trace comparison, the goal is that they conform to their thresholds. When they do not, an operator needs to be informed to take appropriate action before performing new movements of the crane. From this description, it becomes clear that this is a type of watchdog, as was previously discussed. We see that there is no actuation conflict, since the output of the validation is merely used to inform an operator. We are therefore dealing with a \textsf{New Output} transformation, as shown previously in \cref{fig:schema-new-output}. The only difference is that the new output does not go to the \AT, but rather to the operator. Implementation-wise,  the \textsf{Validation} \DT takes as inputs the measured motor state and swing state, as well as the trajectory being currently executed, and compares those measurements with simulations using the model of the gantry crane. When thresholds are breached, the \textsf{Validation} \DT informs the operator of this fact, allowing them to take appropriate action. This is shown in \cref{fig:gantry-crane-evolution2}.

\subsection{Evolution 3: Container Constraints}
\label{sec:containerconstraints}

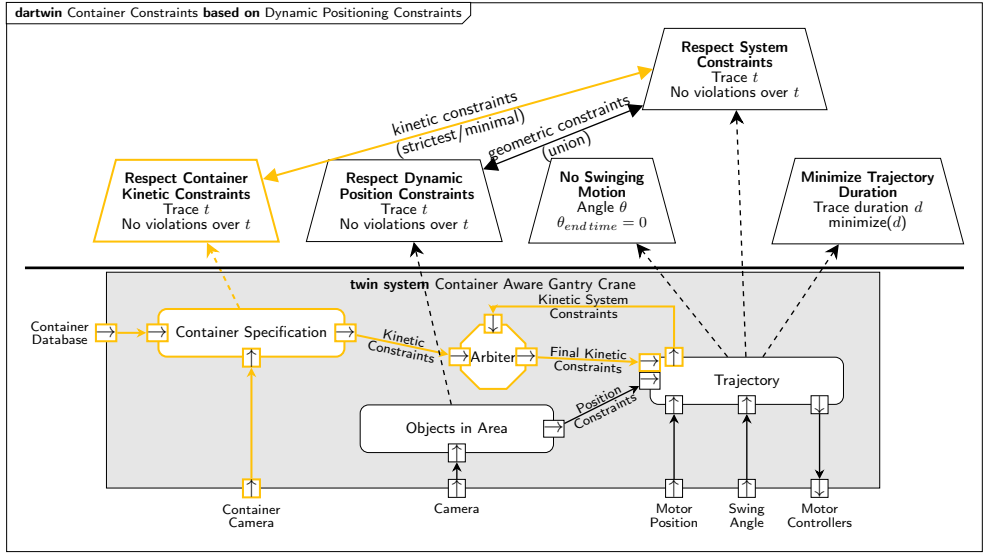
\begin{figure}[htb]
\centering
\resizebox{\textwidth}{!}{
\begin{tikzpicture}[dartwinFramed={Container Constraints}{Dynamic Positioning Constraints}]

\pic {constraints goal};
\pic [below left=of constraints goal] {noswing goal};
\pic [below right=of constraints goal] {minimize goal};
\pic [left=of noswing goal] {nocollision goal};
\draw[goalrelation] (nocollision goal) -- (constraints goal) node[midway,sloped] {geometric constraints\\(union)};

\pic [left=of nocollision goal] {container goal={highlight}};
\draw[goalrelation,highlight] (container goal) -- (constraints goal.west) node[midway,sloped,midway] {kinetic constraints\\(strictest/minimal)};

\begin{scope}[yshift=-7cm]  
    \pic (trajectory-dt) at (.25,.5) {trajectory dt};
    \begin{scope}[on behind layer]
       \node[sys,fill=gray!20,label={[below]north:{\scriptsize \textbf{twin system} Container Aware Gantry Crane}},minimum width=16cm,minimum height=4.5cm] (system) at (-5,0.25) {};
    \end{scope}
    \pic (trajectory-wrap) {wrap trajectory ports={system}};

    \pic (objectsarea-dt) at (-5.75,-.5) {objectsarea dt};
    \node[port,label={[portlabel]below:{{Camera}}}] (system-camera) at (system.south -| objectsarea-dt-camera) {$\uparrow$};
    \draw[flow] (system-camera) -- (objectsarea-dt-camera);

    \pic (container-dt) at (-10,1.5) {container dt={highlight}};
    \pic (container-wrap) {wrap container ports={system}};

    \pic (arbiter-dt) at (-5,1) {gantry arbiter dt={highlight}};

    \node[port,highlight,label={[portlabel]above:{}}] (trajectory-arbiter-out) at ($(trajectory-dt-dt.north)!.75!(trajectory-dt-dt.north west)$) {$\uparrow$};
   
\end{scope}

\node[port,label={below:{\scriptsize }}] (constraints) at (trajectory-dt-dt.west) {$\rightarrow$};
\draw[flow] (objectsarea-dt-constraints) -- node[portlabel,midway,sloped] {Position\\Constraints}(constraints);

\node[port,highlight,label={[align=center]below:{}},above=0 of constraints] (trajectory-dt-limits) {$\rightarrow$};
\draw[flow,highlight] (container-dt-limits) -- node[portlabel,midway,sloped] {Kinetic\\Constraints} (arbiter-dt-west-in);
\draw[flow,highlight] (arbiter-dt-east-out) -- node[portlabel,midway,sloped] {Final Kinetic\\Constraints} (trajectory-dt-limits);

\draw[flow,highlight] (trajectory-arbiter-out) -- ++(0,1.05) -|  node[portlabel,pos=.25,sloped] {Kinetic System\\Constraints} (arbiter-dt-north-in);

\draw[relation] (trajectory-dt-dt) -- (constraints goal);
\draw[relation] (trajectory-dt-dt) -- (noswing goal);
\draw[relation] (trajectory-dt-dt) -- (minimize goal);
\draw[relation] (objectsarea-dt-dt) -- (nocollision goal);
\draw[relation,highlight] (container-dt-dt) -- (container goal);

\pic {frameSeparation={container goal}};  
\end{tikzpicture}
}
\caption{DarTwin Container Constraints adds the \textsf{Respect Container Kinetic Constraints} goal, which is fulfilled by the \textbf{Container Specification} \DT with the help of the \textsf{Arbiter}.}
\label{fig:gantry-crane-evolution4}
\end{figure}

This last evolution is that of handling dynamic container constraints. Perhaps not every container can be moved at the same speed and with the same acceleration due to the contents of that container.
The new goal is thus to have dynamic kinetics constraints dependent on the container that's being moved. The \Ac{poi} are the kinetics constraints of the container, the goal is to adhere to these constraints. Similarly to \cref{sec:dynamicpositioningconstraints}, there is a link to the goal of the \textsf{Trajectory} \DT that needs to ensure all system constraints are met. In contrast to the geometric constraints, which become the union of both, the kinetic constraints the \textsf{Trajectory}  \DT must apply is the strictest combination of both the container and the system constraints. 
At first glance, we could apply the \textsf{Augmented} transformation again, as we did previously in \cref{sec:dynamicpositioningconstraints}. However, we can also apply an \textsf{Arbiter} such as in ~\cref{fig:schema-arbiter}. Indeed, instead of having the \textsf{Trajectory} \DT deciding on which constraints to apply, an \textsf{Arbiter} can do it in its place. To apply the arbiter, the \textsf{Trajectory} \DT must provide the \textsf{Arbiter} with its constraints, and a new \textsf{Container Specification} \DT must provide the \textsf{Arbiter} with the container constraints. The \textsf{Arbiter} then provides the \textsf{Trajectory} \DT with the applicable constraints.
Therefore, to attain this new goal, we introduce a new \textsf{Container Specification} \DT. It connects to a container camera to identify the container being moved, and looks in a container database to find the constraints of the container. It forwards the constraints to the \textsf{Arbiter}, which provides the \textsf{Trajectory} \DT with the strictest set of constraints to apply. The \textsf{Trajectory} \DT must be mutated such that it can provide the \textsf{Arbiter} with its own constraints, and such that it can accept the kinetic constraints. This is all shown in \cref{fig:gantry-crane-evolution4}, where for presentation purposes we chose to take the \textsf{Dynamic Positioning Constraints} DarTwin as basis rather than the \textsf{Continuous Validation} DarTwin.

\subsection{Note on Implementation}

It is worth noting which of the previously discussed elements are actually implemented. The \textsf{Optimal Control} twin system in combination with the \textsf{Continuous Validation} \DT are actually up and running. The crane can perform these optimal moves, and each trajectory is compared to simulation to validate that the crane is operating normally. Providing the dynamic constraints to the \textsf{Trajectory} \DT is also possible. What is not implemented are the \textsf{Objects in Area} \DT and the \textsf{Container Specification} \DT in combination with its \textsf{Arbiter}. For now, constraints imposed by those \DT's have to be defined manually. Implementing those two \DT's would not be unreasonably hard, the container specifications would simply have to be stored and retrieved in/from a database. Somewhat harder are the dynamic keep-out zone classification and segmentation of the \textsf{Objects in Area} \DT, for which deep learning techniques are appropriate.

\section{Discussion}\label{sec:discussion}

In our study, we inferred the different architectural choices without analysing the trade-offs between alternatives. In some situations, several options were applicable. Our preferred choice intended to separate the concerns of the twins as much as possible, as suggested by~\citep{Parnas_separation}. Conversely, we could have always tried to evolve a single twin to encompass all the different concerns. 
Nonetheless, other competing choices might still have been available. To help break the tie in these situations, several antipatterns might be available to help make better decisions~\citep{Ran_anti_patterns}. Another strategy could have been to calculate a set of architectural metrics to help make the decision~\citep{Nakamura_metrics}. Furthermore, developers under release pressure could also calculate technical debt metrics to help make decisions to allow for some technical debt in favour of release time~\citep{nord2012search}.

The presented set of transformations was yielded by our running case study. Typically, every evolution was triggered by some emerging new purpose. Our focus on separation of concerns led to a strategy to investigate whether a simple architectural transformation would be sufficient such that the new goal could be satisfied with as little dependence between the old and the new \DT's. The hierarchical transformation considered the basic \textsf{Thermostat} as is, and just added a new DT that manipulated the input parameters of the existing DT. Then we showed that we could flatten a hierarchical twin system and return with a flattened twin system rather than a hierarchical twin system. Semantically, these architectures are equivalent, but one can easily imagine that extra-functional properties such as the physical possibility to control input parameters may have favoured one of those architectures.

The orthogonal transformation is even more focused on separation of concerns. However, this transformation comes with a requirement, namely that the orthogonal \DTs must not send competing actuations. In our watchdog evolution, the \textsf{Heater} may get conflicting instructions, and we had to abolish the pure orthogonal architecture. Instead, we wanted an architecture which would imply prioritisation between \DTs and thus between their associated goals. In our \textsf{Freeze Protected Thermostat}, we wanted that the \textsf{Freeze Protection} \DT has the priority, and this was achieved by the chained transformation by post-chaining the \textsf{Freeze Protection} \DT. The actuation of the \textsf{Thermostat Logic} \DT is monitored and potentially modified by the \textsf{Freeze Protection} \DT which then forwards the final actuation to the \textsf{Heater}. A post-chained \DT would have priority, while a pre-chained DT would provide input parameters to the sequel DT.

The \textsf{Arbitration} transformation was introduced to respond to the fact that sometimes neither full orthogonality nor priorities are sufficient. There are conflicts between the \DTs that must be harmonized. Logically this means that their corresponding goals must be harmonized and compromised. We have shown two evolutions that apply this transformation. Our example was the \textsf{Compromise Saving} DarTwin, combining both the general \textsf{Lower Energy} goal with the \textsf{Saving Money} goal.

In the \textsf{Gantry Crane} twin system case, we saw the recurrence of four of the different transformations, additionally it also showed how the goals can have relations that are dealt with in different ways. Furthermore, it showed that the notation also works for a system from a different domain and with a different implementation.

In this article, we refer to evolutionary transformations. However, for most software engineers, these transformations are generally referred to as \emph{patterns}: a reusable solution, capturing best practices, to a commonly occurring problem within a given context.  This collection would be referred to as a pattern catalogue, best known from the group-of-four~\citep{gamma1995design}. We, on purpose, do not use the pattern terminology as the evolution of DTs is yet in its infancy and no clear best practices have emerged so far.

\paragraph{Limitations}\label{sec:threats}
Our study is based on a real-world use case that has been actively used in everyday life, thereby 
fulfilling ``the two aspects most clearly distinguishing a case study from other types of research studies are `contemporary phenomenon' and `real-life context'\thinspace''~\citep{WOHLIN2021106514}.
Despite this, however, our work does not claim to be exhaustive or complete, but we instead describe a systematic and careful exploration that hopefully can prove fruitful for future research on \DT evolution. Though an evaluation was performed on a gantry crane case study, a use case from another domain, no new transformations were discovered. Thus, some of our future work involves the discovery and application of evolutionary transformations in/for other case studies and in other domains.

Another dimension that requires further analysis is the extension of our work to (potentially orders of magnitude) larger systems. 
Remaining in the domain of smart buildings, we are planning to expand our private home study to, \eg office buildings.
In this context, we note that our expectation is that the transformations remain similar, however, the types and rate of evolution are expectedly higher.

\section{Related Work}\label{sec:related}
The objective of our research is to provide support for the systematic and iterative development and evolution of \DTs. Our proposed approach focuses on the structural dimension of \DTs and aims at building complex multi-aspects \DTs by composing and evolving simpler ones. In this context, related work includes generic approaches to software evolution, approaches that look at the evolution of \DTs, and the composition of DTs. 

\paragraph{Software Evolution}

Evolution is a natural part of any system's lifecycle. Lehman postulated eight laws of software systems evolution between 1969 and 1996~\citep{lehman1996laws}, with one stating that \emph{``An E-type program must be continually adapted, else it becomes progressively less
satisfactory''}~\citep{lehman1996laws}. This law applies not only to pure software systems but also to software-intensive and cyber-physical systems, \eg in automated production systems~\citep{VOGELHEUSER201554}.
Standards also include the topic of evolution, \eg the ISO25000~\citep{ISO25000} standard on software quality defines four types of important maintenance: 
\begin{inparaenum}[(i)]
    \item corrective maintenance, focusing on resolving issues and errors in the system; 
    \item adaptive maintenance to respond to changes in the environment or requirements of the system; 
    \item perfective maintenance to improve the system (\eg optimise, re-engineer); 
    \item preventive maintenance, focusing on preventing issues before they occur.
\end{inparaenum}
In this article, we mainly concern ourselves with adaptive and perfective maintenance. 

Software evolution is an active research field with various research directions. One such direction is the visualisation of evolutions to understand the system, \eg ~\citep{demeyer2004,pfahler2020visualizing}. However, these visualisations are used to understand the evolution of the software over time and how a change might impact the current twin system. Our approach uses a continuous model that is kept up-to-date and used during the evolution of the twin system. 

Continuous design decision support also has this vision and is close to the work presented here. ~\citep{6617355} presents a documentation model to capture architectural design knowledge during the design and evolution of software systems. This prevents this knowledge and its related assumptions from eroding over time. The model combines elements from both the requirements and architecture. Our notation is similar as it also captures aspects from both the requirements and the architecture side but also contains the properties of interest, which could be conflicting as there is a shared AT. However, our notation is not purely aimed at documentation; it is aimed towards active use during the evolution and also acts as a documentation of the system's current state. 

Design patterns are commonly advocated as a solution for the design and evolution of systems. Durdik created the AM3D (architectural modelling with design decision documentation) pattern catalogue that contains solution-specific questions that can be used to evaluate if a pattern is applicable in a particular case and for documentation~\citep{durdik2016architectural}. The approach also has a workflow that can be adopted. Similarly, we use a pattern-based approach to guide the evolution of our system, where they are reusable solutions to common problems in the evolution of DT-based systems.

\paragraph{Digital Twin Evolution}

Many systematic literature reviews (SLRs) have been published on different aspects of \DTs over the last years \citep{kritzinger2018digital,jones2020characterising, fang2022industry, mohsen2023digital}, but none of them identify papers that directly address the systematic and incremental evolution of \DTs using a transformation approach.

Regarding \DTs evolution, Michael Stych \citep{Arup2019} highlights the successful development of \DTs requires embracing a continuous improvement approach and adopting an iterative approach to identifying and exploring use cases. The development of a \DT for a system of the dimension of a building is unlikely to succeed if it is addressed as a whole all at once. To succeed, the overall building \DT needs to be incrementally and systematically developed, starting from smaller and simpler building components that are individually evolved and assembled. While the paper describes the overall approach, it does not discuss the technical aspects of the incremental development.

In the context of \acfp{cps} where a large amount of data is collected from various sources, \citep{lehner2021towards} proposes the use of a dedicated framework to support the evolution of a \DT in synchrony with the evolution of the system data and data sources it uses as changes are made on the physical system.
The approach aims at reducing the manual effort that is required to synchronise a \DT with the modifications made to a \ac{cps} systems (more precisely to the data and data sources that the \DT is using), ensuring the compatibility of the data schema, and eliminating the unintended loss of data that may occur if synchronisation is not performed correctly. This approach focuses on the data aspect of \DTs, while we focus on the broader perspective of the DTs and the AT.

In the context of smart buildings, \citep{chevallier2020reference} proposes a reference architecture for the development of \DTs that aims at integrating heterogeneous data from different sensor types in a single data repository to enable the development of different applications (which can be viewed as \DT services).
They emphasize the fact that the databases containing building data/information (in this case, the IoT ontology and the sensor database) need to be constantly updated/evolved as sensors and actuators are added/removed/modified in the building to keep the databases in synchrony with the actual building.
In this approach, the evolution of \DTs is related to the evolution of the building, the \AT, and is focused on the data aspect. The concept of \DT evolution is related to the need to maintain the different databases in synchrony with each other in the context of building evolution.
The paper discusses the fact that the integration of all the data in a single repository (that was previously developed in silos) enables the development of different applications to address different aspects of the system, but it does not directly address the question of how the system \DT can be incrementally developed (evolved) to integrate different aspects of a system.

\paragraph{Composition of \DTs}
We consider system evolution an iterative and incremental process. The purpose of the existing \DTs may change over time, while new \DTs for other purposes might be added in other situations. As such, different \DTs have to be combined to focus on specific aspects of improving the system. The idea of composition and interoperability of \DTs aligns with several contributions in literature, although their focus lies on compositional aspects, not evolution. Especially in the case of systems of systems we find references describing how the compositionality of twins is a challenge given conflicting purposes and constraints \citep{michael2022,olsson2023}.

Traore describes the conceptual framework on interoperability between disparate urban \DTs~\citep{traore2024conceptual}. Several levels of interoperability are described: data interoperability, model interoperability, service interoperability, data/model reuse, data/service reuse and model/service reuse. These types of composition are at the technical level and define the space of possible solutions for combining \DTs together. 

Ashtari Talkhestani \etal describe a hierarchical \DT architecture composed of an intelligent layer on top of a set of \DTs~\citep{ashtari2019architecture}. The intelligent \DT layer is responsible for  process optimisation by coordinating the \DTs below. 
Similarly, Reiche et al. propose a networked structure of \DTs based on the system hierarchy~\citep{reiche2021digital}. They introduce a digital twin of a System (DTS) that is superordinate to other DTs. The DTS thus manages the other DTs as an administrator. A DTS2DT-interface component is introduced to provide the communication. The authors focus on the technical aspects of the composition: how to communicate and which information is to be exchanged. Both architectures use the hierarchical transformation to compose the intelligent \DT with the (orthogonal) regular \DT.

Gil at al.\citep{gil2023} describe a DT-based approach to architect cooperative systems. In their approach \DTs have semantic relationships and can be composed in a composition \DT that inherits operations, attributes and behaviours of the composing \DTs. They state this composition of small \DTs may reduce implementation effort in complex systems. They demonstrate their approach by composing the \DT of a cooperative robot cell through the composition of 4 \DTs, on each of the of two heterogeneous robotic arms and the two effectors.
 
\citep{kamel2021digital} explores the adoption of \DTs in the health sector and discusses their potential use for personalised medicine. The paper highlights the fact that different types of \DTs can be developed to address different needs. For example, a DT can be developed for the whole human body, or only a part of a function of it (\eg digestive system), or body organ (\eg liver). It can also be developed for specific diseases or disorders. The paper introduces the concept of \textit{composite digital twins} that allows building more complex DTs by composing/integrating two or more (simpler) DTs. It also introduces the concept of \textit{digital twin levels}, which refers to the level of sophistication of the DT, and \textit{digital twin aggregates}, which are aggregates of DT instances belonging to different individuals, \eg sets covering one family, population group, or the whole population.
Similar ideas are found in the work of Jia \etal, where they propose to create complex \DTs by composing simple ones~\citep{jia2022simple}. An ontology and knowledge graph combines the different aspect \DTs. 
Finally, Robles \etal present an open framework for \DT composition~\citep{robles2023opentwins}. They aim to technically link individual DT entities to create a higher degree DT, allowing knowledge sharing. These types of frameworks can be used as the basis for the technical composition during the evolution proposed in this paper.

\section{Ongoing and Future Work}\label{sec:pfw}
We are in the process of refining the described approach with more analysis techniques while working on applying the proposed approach to additional case studies and on the integration of DevOps principles \citep{kim2021devops} to support continual improvements of \DTs.

In this paper, we illustrate the proposed approach using a smart home system. This case study has allowed establishing the basis of our approach. As part of ongoing work, we are starting to work on several smart building projects. One focuses on the development of a \DT to monitor and control different aspects of climatic rooms at \'{E}cole de technologie sup\'{e}rieure (ETS) in Montr\'{e}al that will be used to support different research projects in construction engineering. The industrial environment provided by these rooms will allow us to validate and evolve different aspects of our approach. We are also working on the development of a \DT to manage different aspects of a research laboratory at ETS, including thermal comfort, air quality, and acoustic comfort. 

In both of these projects, we extend our work to the integration of the DevOps principles of feedback, and continual learning and experimentation to support the evolution and continual improvement of \DTs. Concretely, we are developing the mechanisms to monitor the impact of modifications made on a \DT. This will allow evaluation of whether the modifications are meeting the targeted improvement objectives.

 We can also support the twin system engineers in analysing the notation we proposed. For this, we can draw inspiration from the large body of work that has been done in the systems dynamics community. For example, the causal loop diagram proposed by Forrester~\citep{forrester,Barbrook-Johnson2022} is a notation from systems dynamics. In causal loop diagrams, the connections show causal influence between these different \acp{poi}. These influences can be positive or negative. By using the analysis techniques of system dynamics (given sufficient constraints), positive and negative feedback loops can be identified. As such, they can be used to model the meta-laws or physical laws of the system in an understandable manner. Furthermore, techniques are available to create and validate such causal loop diagrams. This might help designers better understand the dependencies between \acp{poi} and goals. 

One such direction that draws inspiration from this continual feedback from the real world is to go one step further with Forrester System Dynamics, using ``stock and flow diagrams''. Stock and flow diagrams consist of stocks (or levels) and flows (rates) between them. 
A stock is a quantity that could have accumulated in the past. In our example, the temperature is a stock, as it is a reflection of the energy that accumulated. But also the total amount of energy used would be expressed as a stock. 
Flows represent the rate at which the stock is changing at any given instant. Flows either go into a stock (increasing the stock) or flow out of a stock (decreasing the stock). Stocks can connect from and to (infinite) sources, or between stocks. The rate of a flow can be influenced by converters. The semantic domain of the model is a set of differential equations. As such, stock and flow diagrams can be simulated and thus be used to evaluate the impact of a set of choices

\section{Conclusion}\label{sec:conclution}

In recent years, the MBSE community started exploring the impact of system, data and purpose evolution on DTs. 
We recognize that \DTs by their nature have a very close relationship to a dynamic reality (the \AT) and therefore need systematic, timely and reliable evolution. In contrast to evaluating \DTs as individual units, our research specifically tackles the compositional aspects of \DTs and their interplay during continuous and omnipresent system evolution.

In this paper, we investigate the evolution of a set of \DTs using a case study of a real-world smart home system. Starting from a \DT controlling the thermal comfort of a room, different orthogonal and non-orthogonal goals are added to the system. We also showed that the system's properties of interest can interact with each other, resulting in unpredictable effects on the system's functionality and performance if not adequately considered. 
To reason over the impact of these added goals and properties of interest, we provide a notation that helps connect the evolution's \emph{why}, \emph{what} and \emph{how}; connecting the goals, properties of interest and architecture. This notation allows us to reason on valuable transformations at the level of the twin system's architecture. In the course of our study, we identify several architectural transformations, which we describe alongside the system descriptions. Specifically, we identified the following transformations: Basic , Hierarchical, Augmented, Orthogonal, New output, Chained, Arbitration. Furthermore, we discussed how the notation is valuable in evolving the different aspects of a twin system. 
We expect to enhance the notation and progress the research on reasoning about the continuous evolution of \DT-enabled systems based on this paper

\section*{Declarations}
\subsection*{Funding}
Joost Mertens is funded by the Research Foundation - Flanders (FWO) through strategic basic research grant 1SD3421N.

\bibliography{bibliography}

\end{document}